\documentclass[%
 aip,
 jap,
 amsmath,amssymb,
 reprint,%
]{revtex4-1}

\usepackage{graphicx}
\usepackage{dcolumn}
\usepackage{bm}

\usepackage[utf8]{inputenc}
\usepackage[T1]{fontenc}
\usepackage{mathptmx}

\begin{document}

\preprint{AIP/123-QED}

\title{Semilocal exchange-correlation potentials for solid-state calculations:
Current status and future directions}

\author{Fabien Tran}
\affiliation{Institute of Materials Chemistry, Vienna University of Technology,
Getreidemarkt 9/165-TC, A-1060 Vienna, Austria}
\author{Jan Doumont}
\affiliation{Institute of Materials Chemistry, Vienna University of Technology,
Getreidemarkt 9/165-TC, A-1060 Vienna, Austria}
\author{Leila Kalantari}
\affiliation{Institute of Materials Chemistry, Vienna University of Technology,
Getreidemarkt 9/165-TC, A-1060 Vienna, Austria}
\author{Ahmad W. Huran}
\affiliation{Institut f\"{u}r Physik, Martin-Luther-Universit\"{a}t Halle-Wittenberg,
D-06099 Halle, Germany}
\author{Miguel A. L. Marques}
\affiliation{Institut f\"{u}r Physik, Martin-Luther-Universit\"{a}t Halle-Wittenberg,
D-06099 Halle, Germany}
\author{Peter Blaha}
\affiliation{Institute of Materials Chemistry, Vienna University of Technology,
Getreidemarkt 9/165-TC, A-1060 Vienna, Austria}

\date{\today}

\begin{abstract}

Kohn-Sham (KS) density functional theory (DFT) is a very efficient method for
calculating various properties of solids as, for instance, the total energy,
the electron density, or the electronic band structure. The KS-DFT method leads
to rather fast calculations, however the accuracy depends crucially on the
chosen approximation for the exchange and correlation (xc) functional $E_{\text{xc}}$
and/or potential $v_{\text{xc}}$. Here, an overview of xc methods to calculate the
electronic band structure is given, with the focus on the so-called semilocal
methods that are the fastest in KS-DFT and allow to treat systems containing up
to thousands of atoms. Among them, there is the modified Becke-Johnson
potential that is widely used to calculate the fundamental band gap of
semiconductors and insulators. The accuracy for other properties like the
magnetic moment or the electron density, that are also determined directly by
$v_{\text{xc}}$, is also discussed.

\end{abstract}

\maketitle

\section{\label{sec:introduction}Introduction}

The calculation of the properties of solids can be done very efficiently
with the Kohn-Sham\cite{KohnPR65} (KS) method of density functional
theory\cite{HohenbergPR64} (DFT). KS-DFT is clearly much
faster than other methods, like for instance $GW$,\cite{HedinPR65,AryasetiawanRPP98}
that are commonly used for calculating the quasiparticle band structure, or the random-phase
approximation (RPA) and beyond\cite{RenPRB13,ChenARPC17} for the total energy.
Therefore, KS-DFT calculations on solids with a unit cell containing up to several
thousands of atoms can be afforded.
However, in KS-DFT a component in the total energy expression and in the
effective potential in the corresponding KS equations, the one accounting for the
exchange (x) and correlation (c) effects, has to be approximated. Exchange is
due to the Pauli exclusion principle, while correlation arises due to a lowering
of the energy by considering wavefunctions that are
beyond the single determinant approximation in the Hartree-Fock theory.
Since the reliability of the results obtained with KS-DFT depends mainly on the
chosen xc functional, the search for more accurate approximate functionals is a very
active field of research\cite{KuemmelRMP08,CohenCR12,MardirossianMP17}
and several hundreds have been proposed so far.\cite{LehtolaSX18,MardirossianMP17}
Most xc-functionals can be classified on the Jacob's ladder of
John Perdew,\cite{PerdewAIP01,PerdewJCP05}
where the first rung represents the most simple type of approximations
and the highest rung the most sophisticated functionals.
The rather general rule is that the functionals should, in principle, be more accurate,
but also (this is the downside) computationally more expensive to evaluate
when climbing up Jacob's ladder.

The properties of solids which are calculated directly from total energy are for
instance the equilibrium geometry, the cohesive energy, or the formation
enthalpy. For such calculations, a couple of xc functionals
\cite{PerdewPRB92b,PerdewPRL96,HeydJCP03,WuPRB06,PerdewPRL08,SunPRL15}
were or are considered as standard.
Other properties depend more directly on the accuracy
of the xc potential in the KS equations, such as the electronic band structure,
the electron density, and magnetism, for which it may sometimes be
necessary to use other types of xc approximations
\cite{AnisimovPRB91,HeydJCP03,TranPRL09,KuismaPRB10} to get reliable results.

Here, the focus will be on the calculation of the electronic band structure
of semiconductors and insulators with xc methods that are computationally
the fastest, namely the so-called semilocal methods. Needless to say that
quantities derived from the electronic band structure such as the band gap or
the optical constants can serve as a guide for the design of more efficient
materials for technological
applications like photovoltaics,\cite{YuPRL12,OngSR19} light emitting
diodes,\cite{XiaoNP17,MitchellJAP18}
dynamic random access memory,\cite{TraversaN14}
or thermoelectricity.\cite{UrbanJAP19}
Among the semilocal methods,
the modified Becke-Johnson potential proposed by two of
us\cite{TranPRL09} has been shown to be very accurate for the calculation of
the fundamental band gap. Comparison studies for the band gap
(see Refs.~\onlinecite{LeePRB16,TranJPCA17,PandeyJPCC17,NakanoJAP18,TranPRM18,BorlidoJCTC19}
for the most exhaustive comparisons) have shown that the modified Becke-Johnson potential is on
average more accurate than basically all other DFT methods, including the
traditional hybrid functionals using a fixed fraction of Hartree-Fock exchange.
Presently, only dielectric-function-dependent hybrid
functionals\cite{MarquesPRB11,KollerJPCM13,SkonePRB14,ChenPRM18}
and $GW$ methods, in particular if applied self-consistently
including vertex corrections,\cite{ShishkinPRL07,ChenPRB15} can lead to higher accuracy.
However, since the modified Becke-Johnson potential is a DFT method, and furthermore
was constructed empirically, the results are of course not systematically accurate.
In addition, making a xc method accurate for a
specific property (e.g., band gap), may deteriorate the description of
other properties compared to more traditional methods.

The remainder of this paper is organized as follows. In Sec.~\ref{sec:theory}, a brief
introduction to the KS-DFT method is given and the existing
approximations for the xc potential $v_{\text{xc}}$ are reviewed.
Then, Sec.~\ref{sec:overview} provides a summary of the performance of
various potentials for the electronic structure with a particular emphasis
on the fundamental band gap. Results for other properties, e.g., the bandwidth,
the magnetic moment, or the electron density, will also be mentioned.
Finally, a summary is given in Sec.~\ref{sec:summary} and ideas for further
improvements are given in Sec.~\ref{sec:outlook}.

\section{\label{sec:theory}Theoretical background}

\subsection{\label{sec:gKS}(Generalized) Kohn-Sham equations}

In the KS-DFT method,\cite{KohnPR65} the total energy per
unit cell of a periodic solid is given by
\begin{equation}
E_{\text{tot}} =
T_{\text{s}} +
\frac{1}{2}\int\limits_{\text{cell}}
v_{\text{Coul}}(\mathbf{r})\rho(\mathbf{r})d^{3}r -
\frac{1}{2}\sum_{\alpha}^{\text{cell}}Z_{\alpha}
v_{\text{M}}^{\alpha}(\mathbf{R}_{\alpha}) + E_{\text{xc}},
\label{eq:Etot}
\end{equation}
where
\begin{equation}
T_{\text{s}} = -\frac{1}{2}\sum_{\sigma}\sum_{n,\mathbf{k}}
w_{n\mathbf{k}\sigma}\int\limits_{\text{cell}}
\psi_{n\mathbf{k}\sigma}^{*}(\mathbf{r})
\nabla^{2}\psi_{n\mathbf{k}\sigma}(\mathbf{r})d^{3}r
\label{eq:Ts}
\end{equation}
is the noninteracting kinetic energy, the second and third terms
represent the electrostatic interactions
(electron-electron, electron-nucleus, and nucleus-nucleus) with
\begin{equation}
v_{\text{Coul}}(\mathbf{r}) =
\int\limits_{\text{crystal}}\frac{\rho(\mathbf{r}')}
{\left\vert\mathbf{r}-\mathbf{r}'\right\vert}d^{3}r' -
\sum_{\beta}^{\text{crystal}}
\frac{Z_{\beta}}{\left\vert\mathbf{r}-\mathbf{R}_{\beta}\right\vert}
\label{eq:vcoul}
\end{equation}
and
\begin{equation}
v_{\text{M}}^{\alpha}(\mathbf{R}_{\alpha}) =
\int\limits_{\text{crystal}}\frac{\rho(\mathbf{r}')}
{\left\vert \mathbf{R}_{\alpha}-\mathbf{r}'\right\vert}d^{3}r' -
\sum_{\beta\atop\beta\neq\alpha}^{\text{crystal}}
\frac{Z_{\beta}}{\left\vert \mathbf{R}_{\alpha}-\mathbf{R}_{\beta}\right\vert}
\label{eq:vM}
\end{equation}
being the Coulomb and Madelung potentials, respectively,
and $E_{\text{xc}}$ is the exchange-correlation energy.
In Eqs.~(\ref{eq:Etot}), (\ref{eq:vcoul}), and (\ref{eq:vM}),
$\rho=\rho_{\uparrow}+\rho_{\downarrow}$ is the electron density, while $Z_{\alpha}$ and $\mathbf{R}_{\alpha}$
are the charge and position of the nuclei.
In Eq.~(\ref{eq:Ts}), $n$, $\mathbf{k}$, and $\sigma$ are the
band index, $\mathbf{k}$-point, and spin index, respectively,
and $w_{n\mathbf{k}\sigma}$ is the product of the $\mathbf{k}$-point
weight and the occupation number.

Searching for the Slater determinant which minimizes
Eq.~(\ref{eq:Etot}) yields the one-electron Schr\"{o}dinger equations
\begin{equation}
\left(-\frac{1}{2}\nabla^{2} +
v_{\text{Coul}}(\mathbf{r}) +
\hat{v}_{\text{xc},\sigma}(\mathbf{r})\right)\psi_{n\mathbf{k}\sigma}(\mathbf{r}) =
\varepsilon_{n\mathbf{k}\sigma}\psi_{n\mathbf{k}\sigma}(\mathbf{r})
\label{eq:KS}
\end{equation}
that need to be solved self-consistently together with
$\rho_{\sigma}=\sum_{n,\mathbf{k}}w_{n\mathbf{k}\sigma}\left\vert\psi_{n\mathbf{k}\sigma}\right\vert^{2}$.
Within the strict KS framework,
$\hat{v}_{\text{xc},\sigma}$ is calculated as the functional derivative of
$E_{\text{xc}}$ with respect to the electron density $\rho_{\sigma}$
($\hat{v}_{\text{xc},\sigma}=\delta E_{\text{xc}}/\delta\rho_{\sigma}$),
which means that $\hat{v}_{\text{xc},\sigma}$ is a multiplicative potential
($\hat{v}_{\text{xc}}\psi_{n\mathbf{k}\sigma}=v_{\text{xc}}\psi_{n\mathbf{k}\sigma}$),
i.e., it is the same for all orbitals.
Instead, in the generalized KS (gKS) framework,\cite{SeidlPRB96}
the derivative of $E_{\text{xc}}$ is taken with respect to $\psi_{n\mathbf{k}\sigma}^{*}$
($\hat{v}_{\text{xc}}\psi_{n\mathbf{k}\sigma}=
\delta E_{\text{xc}}/\delta\psi_{n\mathbf{k}\sigma}^{*}$),
which leads to a non-multiplicative xc operator (i.e., a potential that is
different for each orbital $\psi_{n\mathbf{k}\sigma}$) in the case of
functionals $E_{\text{xc}}$ that do not depend only explicitly on
$\rho_{\sigma}$.

Functionals whose $\rho_{\sigma}$-dependency is fully explicit are, for instance,
the local density approximation (LDA)\cite{KohnPR65} and generalized
gradient approximations (GGA). The most known examples of functionals that
depend implicitly on $\rho_{\sigma}$ are Hartree-Fock (HF) and the
meta-GGAs (MGGA).\cite{DellaSalaIJQC16}

As mentioned above, there are two types of xc potentials: multiplicative and
non-multiplicative. In addition, among the multiplicative potentials
$v_{\text{xc},\sigma}$ there are two distinctive subgroups, namely, those which
are functional derivatives
$v_{\text{xc},\sigma}=\delta E_{\text{xc}}/\delta\rho_{\sigma}$
of an energy functional $E_{\text{xc}}$ and those which are not,
but were modelled directly. Among the non-multiplicative potentials,
the only ones we are aware of that are not obtained as functional derivative
$\hat{v}_{\text{xc}}\psi_{n\mathbf{k}\sigma}=
\delta E_{\text{xc}}/\delta\psi_{n\mathbf{k}\sigma}^{*}$
are the hybrids with a fraction of HF exchange that depends on a property of
the system like the dielectric function (see, e.g.
Refs.~\onlinecite{MarquesPRB11,KollerJPCM13,SkonePRB14}). The two next sections
list some of the potentials which are the most relevant for the present work,
i.e., for band gaps in particular. Then, in the third section a summary
of what is known about the xc derivative discontinuity, which is strongly related to
the band gap, is provided. Actually, we mention that we are here concerned
with calculated band gaps that are obtained with the orbital energies,
see Sec.~\ref{sec:discontinuity} for more discussion.

\subsection{\label{sec:multv}Multiplicative potentials}

\subsubsection{\label{sec:mpfd}Potentials that are functional derivatives}

We start by defining the xc-energy density per volume $\varepsilon_{\text{xc}}$,
\begin{equation}
E_{\text{xc}} =
\int\limits_{\text{cell}}\varepsilon_{\text{xc}}(\mathbf{r})d^{3}r,
\label{eq:Exc}
\end{equation}
which in LDA is a function of $\rho_{\sigma}$:
\begin{equation}
\varepsilon_{\text{xc}}^{\text{LDA}}(\mathbf{r}) =
\varepsilon_{\text{xc}}^{\text{LDA}}\left(
\rho_{\uparrow}(\mathbf{r}),
\rho_{\downarrow}(\mathbf{r})
\right),
\label{eq:excLDA}
\end{equation}
while in GGA\cite{BeckePRA88,PerdewPRL96} the first derivative of
$\rho_{\sigma}$ is also used:
\begin{equation}
\varepsilon_{\text{xc}}^{\text{GGA}}(\mathbf{r}) =
\varepsilon_{\text{xc}}^{\text{GGA}}
\left(
\rho_{\uparrow}(\mathbf{r}),
\rho_{\downarrow}(\mathbf{r}),
\nabla\rho_{\uparrow}(\mathbf{r}),
\nabla\rho_{\downarrow}(\mathbf{r})
\right).
\label{eq:excGGA}
\end{equation}
The functional derivative of LDA and GGA functionals is straightforwardly calculated
with \cite{Parr}
\begin{equation}
v_{\text{xc},\sigma}^{\text{LDA}} =
\frac{\delta E_{\text{xc}}^{\text{LDA}}}{\delta\rho_{\sigma}} =
\frac{\partial \varepsilon_{\text{xc}}^{\text{LDA}}}{\partial\rho_{\sigma}}
\label{eq:dExcLDA}
\end{equation}
and
\begin{equation}
v_{\text{xc},\sigma}^{\text{GGA}} =
\frac{\delta E_{\text{xc}}^{\text{GGA}}}{\delta\rho_{\sigma}} =
\frac{\partial \varepsilon_{\text{xc}}^{\text{GGA}}}{\partial\rho_{\sigma}} -
\nabla\cdot\left(\frac{\partial\varepsilon_{\text{xc}}^{\text{GGA}}}{\partial\nabla\rho_{\sigma}}\right),
\label{eq:dExcGGA}
\end{equation}
respectively. From Eqs.~(\ref{eq:dExcLDA}) and (\ref{eq:dExcGGA}),
we can see that a GGA potential depends on $\rho_{\sigma}$ and the first and
second derivatives of $\rho_{\sigma}$, while $v_{\text{xc},\sigma}^{\text{LDA}}$ depends
only on $\rho_{\sigma}$.

Among the functionals $E_{\text{xc}}$ of the LDA type that will be considered for the
discussion in Sec.~\ref{sec:overview}, there is the functional of the
homogeneous electron gas\cite{KohnPR65,VoskoCJP80,PerdewPRB92a} (called LDA)
and Sloc\cite{FinzelIJQC17} (local Slater potential), which is
an enhanced exchange LDA with no correlation and was proposed specifically
for band gap calculations.

With something like 200 functionals, the GGAs represent the largest group
of functionals,\cite{LehtolaSX18} however only a couple of them were shown
to be interesting for the band gap. Among them, the three following are
here considered. EV93PW91, which consists of the
exchange EV93 of Engel and Vosko\cite{EngelPRB93} that we combined with PW91
correlation\cite{PerdewPRB92b} in our previous works.\cite{TranJPCM07,TranJPCA17}
The EV93 exchange was constructed to reproduce the exact exchange (EXX) potential
in atoms. AK13 from Armiento and K\"{u}mmel,\cite{ArmientoPRL13} a
parameter-free exchange functional that was constructed
to have a potential that changes discontinuously at integer
particle numbers. In our previous works,\cite{TranJPCA17}
as well as in others,\cite{VlcekPRB15} AK13 was used with no correlation added.
In the recently proposed GGA HLE16,\cite{VermaJPCL17}
the parameters were tuned in order to give accurate band gaps.
Of course, the standard GGA for solids of Perdew
\textit{et al}.\cite{PerdewPRL96} (PBE),
which is known to be very inaccurate for band gaps,\cite{HeydJCP05}
will also be considered [note that a simple relationship between PBE
and one-shot $GW$ ($G_{0}W_{0}$) band gaps was proposed in
Ref.~\onlinecite{MoralesGarciaJPCC17}]. Other standard GGAs like
BLYP\cite{BeckePRA88,LeePRB88} or PBEsol\cite{PerdewPRL08} lead
to results for the electronic structure that are very similar to
PBE\cite{TranPRB15} and, therefore, do not need to be considered.

The other families of pure DFT approximations for $E_{\text{xc}}$
like the $\nabla^{2}\rho$-MGGA
functionals\cite{DellaSalaIJQC16} which depend on the second derivative of
$\rho$, the nonlocal van der Waals functionals,\cite{DionPRL04} or the
weighted-density approximation \cite{GunnarssonSSC77,AlonsoPRB78}
are not considered in the present work. We just mention that the latter two
approximations are more complicated since $\varepsilon_{\text{xc}}$
is itself an integral:
\begin{equation}
\varepsilon_{\text{xc}}(\mathbf{r}) = \int\limits_{\text{crystal}}
f\left(\rho(\mathbf{r}),\rho(\mathbf{r}'),
\nabla\rho(\mathbf{r}),\nabla\rho(\mathbf{r}'),
\left\vert\mathbf{r}-\mathbf{r}'\right\vert\right)
d^{3}r',
\label{eq:excnl}
\end{equation}
which brings full nonlocality (they are beyond the semilocal approximations),
but also leads to more complicated implementations
and expensive calculations.

\subsubsection{\label{sec:mpnfd}Potentials that are not functional derivatives}

A couple of potentials $v_{\text{xc}}$ were directly modelled in order
to produce accurate results for a given property, typically related to the
electronic structure, like the band gap. Such potentials having no associated
energy functional $E_{\text{xc}}$ were named
\textit{stray} by Gaiduk \textit{et al}.\cite{GaidukJCTC09}

Presenting such potentials chronologically, we start with LB94,\cite{vanLeeuwenPRA94}
which reads
\begin{equation}
v_{\text{xc},\sigma}^{\text{LB94}}(\mathbf{r}) =
v_{\text{xc},\sigma}^{\text{LDA}}(\mathbf{r}) -
\beta\rho_{\sigma}^{1/3}(\mathbf{r})\frac{y_{\sigma}^{2}(\mathbf{r})}{1+3\beta y_{\sigma}(\mathbf{r})
\text{arcsinh}(y_{\sigma}(\mathbf{r}))},
\label{eq:LB94}
\end{equation}
where $y_{\sigma}=\left\vert\nabla\rho_{\sigma}\right\vert/\rho_{\sigma}^{4/3}$
and $\beta=0.05$. LB94 was constructed such that the asymptotic behavior at
$\left\vert\mathbf{r}\right\vert\rightarrow\infty$ in finite systems is
$-1/\left\vert\mathbf{r}\right\vert$ as it should be.
Note that in Ref.~\onlinecite{SchipperJCP00}, a slight modification of
LB94 (LB94$\alpha$) was proposed, where $\beta=0.01$ and the exchange
$v_{\text{x},\sigma}^{\text{LDA}}$ is multiplied by $\alpha=1.19$.
From Eq.~(\ref{eq:LB94}), we can see that the LB94 potential depends on
the first derivative of $\rho_{\sigma}$, but not on the second derivative
like the GGA potentials do.

Becke and Johnson\cite{BeckeJCP06} (BJ) proposed an approximation to the
EXX potential in atoms that is given by
\begin{equation}
v_{\text{x},\sigma}^{\text{BJ}}(\mathbf{r}) =
v_{\text{x},\sigma}^{\text{BR}}(\mathbf{r}) +
\frac{1}{\pi}\sqrt{\frac{5}{6}}
\sqrt{\frac{t_{\sigma}(\mathbf{r})}{\rho_{\sigma}(\mathbf{r})}},
\label{eq:BJ}
\end{equation}
where
\begin{equation}
v_{\text{x},\sigma}^{\text{BR}}(\mathbf{r}) =
-\frac{1}{b_{\sigma}(\mathbf{r})}\left(1 - e^{-x_{\sigma}(\mathbf{r})} -
\frac{1}{2}x_{\sigma}(\mathbf{r})e^{-x_{\sigma}(\mathbf{r})}\right)
\label{eq:vxBR}
\end{equation}
is the Becke-Roussel (BR) potential\cite{BeckePRA89} with
$x_{\sigma}$ that is obtained by solving a nonlinear equation
involving $\rho_{\sigma}$, $\nabla\rho_{\sigma}$,
$\nabla^{2}\rho_{\sigma}$, and
the kinetic-energy density
$t_{\sigma}=\left(1/2\right)\sum_{n,\mathbf{k}}
w_{n\mathbf{k}\sigma}
\nabla\psi_{n\mathbf{k}\sigma}^{*}\cdot
\nabla\psi_{n\mathbf{k}\sigma}$.
Then, in Eq.~(\ref{eq:vxBR}) $b_{\sigma}=\left[x_{\sigma}^{3}e^{-x_{\sigma}}/
\left(8\pi\rho_{\sigma}\right)\right]^{1/3}$.
Note that in Refs.~\onlinecite{BeckePRA89,BeckeJCP06,TranJCTC15}
the BR potential was shown to reproduce very accurately the
Slater potential,\cite{SlaterPR51} which is the hole component
of the EXX potential. Since the BJ potential depends on $t_{\sigma}$
it can be considered as a MGGA, although it is somehow abusive since
the mathematical structure of $v_{\text{x},\sigma}^{\text{BJ}}$ differs significantly
from the true MGGA potentials discussed in Sec.~\ref{sec:nonmultv}.
The BJ potential has attracted a lot of interest and has been
studied\cite{TranJPCM07,GaidukJCP08,StaroverovJCP08,GaidukJCP09} or modified
\cite{ArmientoPRB08,KarolewskiJCTC09,TranPRL09,RasanenJCP10,PittalisPRB10,TranPRB15}
by a certain number of groups.

In particular, among these variants of the BJ potential there is the
aforementioned mBJLDA xc potential that was introduced in Ref.~\onlinecite{TranPRL09}
as an alternative to the expensive $GW$ and hybrid methods for calculating
band gaps. mBJLDA consists of a modification of the BJ
potential (mBJ) for exchange and LDA\cite{VoskoCJP80,PerdewPRB92a} for the
correlation potential. The mBJ exchange is given by
\begin{equation}
v_{\text{x},\sigma}^{\text{mBJ}}(\mathbf{r}) =
cv_{\text{x},\sigma}^{\text{BR}}(\mathbf{r}) +
\left(3c-2\right)\frac{1}{\pi}\sqrt{\frac{5}{6}}
\sqrt{\frac{t_{\sigma}(\mathbf{r})}{\rho_{\sigma}(\mathbf{r})}},
\label{eq:mBJ}
\end{equation}
where $c$ is a functional of the density and is given by
\begin{equation}
c = \alpha + \beta\sqrt{g},
\label{eq:c}
\end{equation}
where
\begin{equation}
g = \frac{1}{V_{\text{cell}}}\int\limits_{\text{cell}}\frac{1}{2}
\left(\frac{\left\vert\nabla\rho^{\uparrow}(\mathbf{r}')\right\vert}
{\rho^{\uparrow}(\mathbf{r}')}
+\frac{\left\vert\nabla\rho^{\downarrow}(\mathbf{r}')\right\vert}
{\rho^{\downarrow}(\mathbf{r}')}\right)d^{3}r'
\label{eq:g}
\end{equation}
is the average of $\left\vert\nabla\rho_{\sigma}\right\vert/\rho_{\sigma}$
in the unit cell. It is important to underline that using $g$
brings some kind of nonlocality since the value of
$v_{\text{x},\sigma}^{\text{mBJ}}$ at $\mathbf{r}$ depends on
$\left\vert\nabla\rho_{\sigma}\right\vert/\rho_{\sigma}$ at every point
in space (thus, mBJLDA is not strictly speaking a semilocal method).
However, this nonlocality is different from the true
nonlocality as in Eq.~(\ref{eq:excnl}) or in the HF method
(see Sec.~\ref{sec:nmpfd}), where there is a dependency
on the interelectronic distance $\left\vert\mathbf{r}-\mathbf{r}'\right\vert$.
The values of $\alpha$ and $\beta$ in Eq.~(\ref{eq:c}) were determined to be
$\alpha=-0.012$ (dimensionless) and $\beta=1.023$~bohr$^{1/2}$
by minimizing the mean absolute error of the band gap for a group of
solids.\cite{TranPRL09} Subsequently, other parametrizations for
$\alpha$ and $\beta$ have been proposed in Ref.~\onlinecite{KollerPRB12}
for semiconductors with band gaps smaller than 7~eV
or in Refs.~\onlinecite{JishiJPCC14,TraorePRB19} for halide perovskites.
Note that in the literature, the mBJLDA potential is sometimes called
TB-mBJ,\cite{SinghPRB10b} TB,\cite{BartokPRB19} or
TB09,\cite{MarquesPRB11,MarquesCPC12,LehtolaSX18}
which refers to the authors of the method.

A very interesting potential for band gap calculations was proposed by
Kuisma \textit{et al}.\cite{KuismaPRB10} Their potential, which is based on the
potential proposed by Gritsenko \textit{et al}.\cite{GritsenkoPRA95} (GLLB),
is given by (SC stands for solid and correlation)
\begin{eqnarray}
v_{\text{xc},\sigma}^{\text{GLLB-SC}}(\mathbf{r}) & = &
2e_{\text{x},\sigma}^{\text{PBEsol}}(\mathbf{r}) + K_{\text{x}}^{\text{LDA}}\nonumber \\
 & & \times\sum_{n,\mathbf{k}}
w_{n\mathbf{k}\sigma}
\sqrt{\epsilon_{\text{H}}-\epsilon_{n\mathbf{k}\sigma}}
\frac{\left\vert\psi_{n\mathbf{k}\sigma}(\mathbf{r})\right\vert^{2}}{\rho_{\sigma}(\mathbf{r})} +
v_{\text{c},\sigma}^{\text{PBEsol}}(\mathbf{r}),\nonumber \\
\label{eq:vxcGLLBSC}
\end{eqnarray}
where
$e_{\text{x},\sigma}^{\text{PBEsol}}$ is the PBEsol exchange-energy density per
spin-$\sigma$ electron [defined by
$E_{\text{x}}=\sum_{\sigma}\int_{\text{cell}}e_{\text{x},\sigma}(\mathbf{r})\rho_{\sigma}(\mathbf{r})d^{3}r$],
$v_{\text{c},\sigma}^{\text{PBEsol}}=\delta E_{\text{c}}^{\text{PBEsol}}/\delta\rho_{\sigma}$
is the PBEsol correlation potential, and
$K_{\text{x}}^{\text{LDA}}=8\sqrt{2}/\left(3\pi^2\right)$.
The particularity of the GLLB-SC potential is to depend on the
orbital energies [$\epsilon_{\text{H}}$ is for the highest occupied one, i.e., at the
valence band maximum (VBM)]. It is easy to show\cite{BaerendsPCCP17} that the
dependency on $\epsilon_{\text{H}}$ in Eq.~(\ref{eq:vxcGLLBSC}) leads to 
the possibility to calculate the exchange part of the
derivative discontinuity, that is given by
\begin{eqnarray}
\Delta_{\text{x}}^{\text{GLLB-SC}} & = &
\int\limits_{\text{cell}}\psi_{\text{L}}^{*}(\mathbf{r})\left[
\sum_{n,\mathbf{k}}K_{\text{x}}^{\text{LDA}}
w_{n\mathbf{k}\sigma_{\text{L}}}
\left(
\sqrt{\epsilon_{\text{L}}-\epsilon_{n\mathbf{k}\sigma_{\text{L}}}} \right.\right. \nonumber\\
& &
-\left.\left.
\sqrt{\epsilon_{\text{H}}-\epsilon_{n\mathbf{k}\sigma_{\text{L}}}}
\right)
\frac{\left\vert\psi_{n\mathbf{k}\sigma_{\text{L}}}(\mathbf{r})
\right\vert^{2}}{\rho_{\sigma_{\text{L}}}(\mathbf{r})}
\right]\psi_{\text{L}}(\mathbf{r})d^{3}r,
\label{eq:deltax}
\end{eqnarray}
where $\psi_{\text{L}}$ is the lowest unoccupied orbital [conduction
band minimum (CBM)] and $\sigma_{\text{L}}$ its spin value.

Among the other computationally fast model potentials proposed in the literature
that are not considered in the present work, we mention the
exchange potentials of Lembarki \textit{et al}.\cite{LembarkiPRA95}
and Harbola and Sen,\cite{HarbolaJPB02,HarbolaJPB02} which, as LB94, depend
on $\rho_{\sigma}$ and $\nabla\rho_{\sigma}$. 
The exchange potential of Umezawa\cite{UmezawaPRA06} depends on $\rho_{\sigma}$ and
$\nabla\rho_{\sigma}$, but also on the Fermi-Amaldi potential\cite{FermiAIR34}
which depends on the number of electrons $N$ such that its application to solids is
unclear. In Ref.~\onlinecite{FerreiraPRB08}, Ferreira \textit{et al}.
proposed the LDA-1/2 method, which improves upon LDA for band gaps.
However, the method is not always straightforward to apply and ambiguities
may arise depending on the case (see Refs.~\onlinecite{XueCMS18,DoumontPRB19}).
A few other model potentials can be found in the work of
Staroverov.\cite{StaroverovJCP08}

We also mention that stray potentials have undesirable features
both at the fundamental and practical level as shown in
Refs.~\onlinecite{GaidukJCP09,GaidukJCTC09,KarolewskiPRA13}.

Last but not least, we mention that some of the potentials presented
in this section (LB94, BJ, and mBJ) and in the previous one (e.g., AK13)
were shown to lead to numerical problems in the asymptotic region
far from nuclei in finite systems.\cite{AschebrockPRB17a,AschebrockPRB17b}

\subsection{\label{sec:nonmultv}Non-multiplicative potentials}

\subsubsection{\label{sec:nmpfd}Potentials that are functional derivatives}

We begin with the MGGA functionals that use the kinetic-energy density
$t_{\sigma}$ as additional ingredient compared to the GGAs
(here, we do not consider MGGAs which depend on $\nabla^{2}\rho$):
\begin{equation}
\varepsilon_{\text{xc}}^{\text{MGGA}} = \varepsilon_{\text{xc}}^{\text{MGGA}}
\left(
\rho_{\uparrow},\rho_{\downarrow},
\nabla\rho_{\uparrow},\nabla\rho_{\downarrow},
t_{\uparrow},t_{\downarrow}
\right).
\label{eq:excMGGA}
\end{equation}
Since there is, via $t_{\sigma}$, an implicit dependency on $\rho_{\sigma}$,
the functional derivative of $E_{\text{xc}}^{\text{MGGA}}$ with respect to
$\rho_{\sigma}$ can not be calculated (at least not in a straightforward way).
Instead, the functional derivative with respect to $\psi_{n\mathbf{k}\sigma}^{*}$
is taken,\cite{NeumannMP96,ArbuznikovCPL03} which gives
\begin{eqnarray}
\hat{v}_{\text{xc},\sigma}^{\text{MGGA}}\psi_{n\mathbf{k}\sigma} & = &
\frac{\partial\varepsilon_{\text{xc}}^{\text{MGGA}}}{\partial\rho_{\sigma}}\psi_{n\mathbf{k}\sigma} -
\left(\nabla\cdot\frac{\partial\varepsilon_{\text{xc}}^{\text{MGGA}}}{\partial\nabla\rho_{\sigma}}\right)
\psi_{n\mathbf{k}\sigma} \nonumber\\
& &
-\frac{1}{2}\nabla\cdot\left(\frac{\partial\varepsilon_{\text{xc}}^{\text{MGGA}}}{\partial t_{\sigma}}
\nabla\psi_{n\mathbf{k}\sigma}\right) \nonumber\\
& = &
v_{\text{xc},\sigma}^{\text{GGA-part}}\psi_{n\mathbf{k}\sigma} -
\frac{1}{2}\nabla\cdot\left(\frac{\partial\varepsilon_{\text{xc}}^{\text{MGGA}}}{\partial t_{\sigma}}
\nabla\psi_{n\mathbf{k}\sigma}\right), \nonumber\\
\label{eq:vxcMGGA}
\end{eqnarray}
where the last term, which is non-multiplicative, arises due to the
dependency on $t_{\sigma}$. Among the numerous MGGA functionals
that have been proposed,\cite{DellaSalaIJQC16} some of them have been
shown to improve clearly over the standard PBE for the
geometry and energetics of electronic systems.
This includes SCAN\cite{SunPRL15} and MVS\cite{SunPNAS15}
(tested in Refs.~\onlinecite{YangPRB16,JanaJCP18a} for the band gap), as well as
HLE17\cite{VermaJPCC17} and revM06-L.\cite{WangPNAS17}

Another class of functionals that lead to a non-multiplicative potential,
but not of the semilocal type, is HF and the (screened) hybrid functionals, where a fraction
$\alpha_{\text{x}}$ of semilocal exchange is replaced by HF exchange:
\begin{equation}
\varepsilon_{\text{xc}}^{\text{hybrid}}(\mathbf{r}) =
\varepsilon_{\text{xc}}^{\text{SL}}(\mathbf{r}) +
\alpha_{\text{x}}\left[\varepsilon_{\text{x}}^{\text{(scr)HF}}(\mathbf{r}) -
\varepsilon_{\text{x}}^{\text{(scr)SL}}(\mathbf{r})\right],
\label{eq:Exchybrid}
\end{equation}
where
\begin{eqnarray}
\varepsilon_{\text{x}}^{\text{(scr)HF}}(\mathbf{r}) & = &
-\frac{1}{2}\sum_{\sigma}
\sum_{n,\mathbf{k}}
\sum_{n',\mathbf{k}'}
w_{n\mathbf{k}\sigma}w_{n'\mathbf{k}'\sigma}
\psi_{n\mathbf{k}\sigma}^{*}(\mathbf{r})
\psi_{n'\mathbf{k}'\sigma}(\mathbf{r}) \nonumber \\
& & \times\int\limits_{\text{crystal}}
v\left(\left\vert\mathbf{r}-\mathbf{r}'\right\vert\right)
\psi_{n'\mathbf{k}'\sigma}^{*}(\mathbf{r}')
\psi_{n\mathbf{k}\sigma}(\mathbf{r}')
d^{3}r', \nonumber \\
\label{eq:ExHF}
\end{eqnarray}
which is a fully nonlocal exchange-energy density.
The functional derivative of $E_{\text{x}}^{\text{(scr)HF}}$ with
respect to $\psi_{n\mathbf{k}\sigma}^{*}$ is given by
\begin{eqnarray}
\hat{v}_{\text{x},\sigma}^{\text{(scr)HF}}\psi_{n\mathbf{k}\sigma}(\mathbf{r}) & = &
-\sum_{n',\mathbf{k}'}w_{n'\mathbf{k}'\sigma}\psi_{n'\mathbf{k}'\sigma}(\mathbf{r}) \nonumber \\
& & \times 
\int\limits_{\text{crystal}}
v\left(\left\vert\mathbf{r}-\mathbf{r}'\right\vert\right)
\psi_{n'\mathbf{k}'\sigma}^{*}(\mathbf{r'})
\psi_{n\mathbf{k}\sigma}(\mathbf{r'})
d^{3}r'. \nonumber \\
\label{eq:vxHF}
\end{eqnarray}
In Eqs.~(\ref{eq:ExHF}) and (\ref{eq:vxHF}), $v$ can be either the bare
Coulomb potential $v=1/\left\vert\mathbf{r}-\mathbf{r}'\right\vert$ or
a potential that is screened at short or long range.
In the case of solids, it is computationally advantageous to
use a potential that is short range, i.e.,
the long-range part is screened. Such short-range potentials
are the Yukawa potential
$v=e^{-\lambda\left\vert\mathbf{r}-\mathbf{r}'\right\vert}/\left\vert\mathbf{r}-\mathbf{r}'\right\vert$
(Ref.~\onlinecite{BylanderPRB90}) or
$v=\text{erfc}(\mu\left\vert\mathbf{r}-\mathbf{r}'\right\vert)/\left\vert\mathbf{r}-\mathbf{r}'\right\vert$
where erfc is the complementary error function (Ref.~\onlinecite{HeydJCP03}).
Due to the nonlocal character of Eqs.~(\ref{eq:ExHF}) and (\ref{eq:vxHF})
as well as the summations over the occupied orbitals, the HF/hybrid methods
lead to calculations that are one or several orders of magnitude more
expensive that with semilocal approximations. However, similarly as $G_{0}W_{0}$,
hybrids can be applied non-self-consistently, which can be a very good
approximation for the band gap.\cite{AlkauskasPB07,TranPLA12}
The most popular hybrid functional in solid-state physics is HSE06
from Heyd \textit{et al}.,\cite{HeydJCP03,KrukauJCP06} which uses
the Coulomb operator screened with the erfc function and the
PBE functional for the semilocal part in Eq.~(\ref{eq:Exchybrid}).
Another well-known hybrid functional is B3PW91,\cite{BeckeJCP93b}
which has been shown to perform very well for band gaps
similarly as HSE06.\cite{CrowleyJPCL16,TranJPCA17}

Another type of non-multiplicative potentials that are not considered here
are those of the self-interaction corrected functionals.\cite{PerdewPRB81}
We also note that the DFT+$U$\cite{AnisimovPRB91} and
onsite-hybrid\cite{NovakPSSB06,TranPRB06} methods also lead to
non-multiplicative potentials, however they are somehow crude
approximations to the HF/hybrid methods.

We mention that in principle the functional
derivative with respect to $\rho$ can also be calculated for implicit
functionals of $\rho$. However, the equations of the optimized effective
potential\cite{SharpPR53,TalmanPRA76,KuemmelRMP08,Engel} (OEP) need to be solved, which,
depending on the type of basis set, can be quite cumbersome
(see, e.g., Ref.~\onlinecite{BetzingerPRB12}) in particular since
the response function, that involves the unoccupied orbitals, is required.

To finish, we also mention that MGGA functionals can be \textit{deorbitalized}
by replacing $t_{\sigma}$ in Eq.~(\ref{eq:excMGGA}) by an approximate expression
that depends on $\rho_{\sigma}$ and its two first
derivatives.\cite{MejiaRodriguezPRA17,MejiaRodriguezPRB18,BienvenuJCTC18,TranJCP18}
This leads to $\nabla^{2}\rho$-MGGAs that are pure DFT functionals with
a potential $v_{\text{xc}}$ that can be readily calculated, but that involves
up to the fourth derivative of $\rho_{\sigma}$.

\subsubsection{\label{sec:nmpnfd}Potentials that are not functional derivatives}

Non-multiplicative potentials that are not obtained as a functional derivative,
but were directly modelled are the hybrids with a fraction $\alpha_{\text{x}}$
of HF exchange that depends on a property of the system like the dielectric function
\cite{AlkauskasPRL08,ShimazakiCPL08,MarquesPRB11,KollerJPCM13,SkonePRB14,GerosaPRB15,CuiJPCL18,ChenPRM18}
or the electron density like in Eq.~(\ref{eq:g}).
\cite{MarquesPRB11,KollerJPCM13}
To our knowledge, no non-multiplicative potential of the semilocal type has
been proposed.

\subsection{\label{sec:discontinuity}Derivative discontinuity}

In KS-DFT,
\begin{equation}
E_{\text{g}}^{\text{KS}}=\varepsilon_{\text{L}}-\varepsilon_{\text{H}}
\label{eq:Eg1}
\end{equation}
is called the KS band gap, which, in \textit{exact} KS-DFT, is not equal to
the fundamental band gap defined as ($N$ is the number of electrons in the system)
\begin{eqnarray}
E_{\text{g}} & = & I(N) - A(N) \nonumber \\
& = & [E_{\text{tot}}(N-1)-E_{\text{tot}}(N)] -
[E_{\text{tot}}(N)-E_{\text{tot}}(N+1)], \nonumber\\
\label{eq:Eg2}
\end{eqnarray}
where $I$ and $A$ are the ionization potential and electron affinity, respectively.
The two gaps differ by the so-called
xc derivative discontinuity $\Delta_{\text{xc}}$:\cite{PerdewPRL82,ShamPRL83}
\begin{eqnarray}
E_{\text{g}} & = & I(N) - A(N) = - \epsilon_{\text{H}}(N) - [-\epsilon_{\text{H}}(N+1)] \nonumber \\
& = & \underbrace{\epsilon_{\text{L}}(N) - \epsilon_{\text{H}}(N)}_{E_{\text{g}}^{\text{KS}}} +
\underbrace{\epsilon_{\text{H}} (N+1) - \epsilon_{\text{L}} (N)}_{\Delta_{\text{xc}}} =
E_{\text{g}}^{\text{KS}} + \Delta_{\text{xc}}, \nonumber\\
\label{eq:deltaxc} 
\end{eqnarray}
where we have used the fact that $I(N)=-\epsilon_{\text{H}}(N)$\cite{LevyPRA84}
in exact KS-DFT.
$\Delta_{\text{xc}}$ is positive and
numerical examples (see Refs.~\onlinecite{GruningJCP06,GruningPRB06,KlimesJCP14}
for results on solids) strongly indicate that $\Delta_{\text{xc}}$ can be of
the same order of magnitude as $E_{\text{g}}^{\text{KS}}$, i.e., the exact
$E_{\text{g}}^{\text{KS}}$ is much smaller than the experimental band gap
$E_{\text{g}}=E_{\text{g}}^{\text{expt}}$.

Concerning approximate xc methods, a brief summary of some of the most
important points from Refs.~\onlinecite{KuemmelRMP08,YangJCP12,PerdewPNAS17,BaerendsJCP18}
is the following:
\begin{itemize}
\item LDA, GGA:
Like with exact KS-DFT, using the orbital energies [Eq.~(\ref{eq:Eg1})]
(i.e., ignoring $\Delta_{\text{xc}}$) leads to band gaps that are much smaller
than experiment with most LDA/GGA functionals for both finite
systems and solids. For finite systems, using the total energies [Eq.~(\ref{eq:Eg2})]
or methods\cite{AndradePRL11,ChaiPRL13,KraislerJCP14,GorlingPRB15}
to calculate $\Delta_{\text{xc}}$ with the LDA/GGA quantities (and add it to
$E_{\text{g}}^{\text{KS}}$) gives much better agreement with experiment.
According to our knowledge, for solids there is no method providing a
non-zero $\Delta_{\text{xc}}$ within a LDA/GGA framework.
Actually, in Ref.~\onlinecite{GorlingPRB15} it is shown that
$\Delta_{\text{xc}}=E_{\text{g}}-E_{\text{g}}^{\text{KS}}=0$ in solids
(see also Ref.~\onlinecite{KraislerJCP14}),
which also means that using Eq.~(\ref{eq:Eg2}) does not help.
Specialized GGA functionals\cite{ArmientoPRL13,VermaJPCL17} can lead to
band gaps calculated with Eq.~(\ref{eq:Eg1}) that agree quite well with experiment.
\item Non-multiplicative potentials:
Potentials of the MGGA and HF/hybrid functionals implemented within the gKS method
lead to a gKS gap $E_{\text{g}}^{\text{gKS}}=\varepsilon_{\text{L}}-\varepsilon_{\text{H}}$
that contains a portion of $\Delta_{\text{xc}}$. For instance, with hybrid functionals
[Eq.~(\ref{eq:Exchybrid})], a fraction $\alpha_{\text{x}}$ of $\Delta_{\text{x}}$
is included in $E_{\text{g}}^{\text{gKS}}$. Thus, this is one of the reasons why
$E_{\text{g}}^{\text{gKS}}$ can be larger (and in better agreement with
experiment) than $E_{\text{g}}^{\text{KS}}$ calculated from standard
LDA/GGA multiplicative potentials.
\item Multiplicative potentials obtained from the OEP method:
For a given non-pure-DFT functional (MGGA, HF/hybrid, or RPA),
the multiplicative potential leads to a KS gap $E_{\text{g}}^{\text{KS}}$ that
is usually smaller than its counterpart $E_{\text{g}}^{\text{gKS}}$, since
$E_{\text{g}}^{\text{gKS}}-E_{\text{g}}^{\text{KS}}\approx\Delta_{\text{xc}}$
that should be positive in principle. However, with MGGA functionals a negative
$\Delta_{\text{xc}}$ can sometimes be obtained.\cite{EichJCP14,YangPRB16}
\end{itemize}

From above, an important point is that it is in principle not correct
to compare $E_{\text{g}}^{\text{KS}}$ with the experimental
$E_{\text{g}}^{\text{expt}}$ when the orbital
energies were obtained with a multiplicative potential. With such potentials,
a $\Delta_{\text{xc}}$ should be added to $E_{\text{g}}^{\text{KS}}$ before
comparing with $E_{\text{g}}^{\text{expt}}$ (however with LDA/GGA for solids
one may argue that it is not necessary since $\Delta_{\text{xc}}=0$).
Theoretically, it is also not very sound to devise multiplicative potentials
that give $E_{\text{g}}^{\text{KS}}$ in agreement with
$E_{\text{g}}^{\text{expt}}$ (mBJLDA and HLE16 are such examples),
since it means that these potentials will most likely be very different from the
exact multiplicative potential. Forcing an agreement between
$E_{\text{g}}^{\text{KS}}$ and $E_{\text{g}}^{\text{expt}}$
may result in a potential with unphysical features leading to quite
inaccurate results for properties other than the band gap, as shown in
Refs.~\onlinecite{WaroquiersPRB13,TranPRM18,TraorePRB19}.

Thus, among the fast semilocal methods, the GLLB-SC potential
($\Delta_{\text{xc}}$ can be calculated) and the
non-multiplicative MGGA potentials ($\Delta_{\text{xc}}$ included in
$\varepsilon_{\text{L}}-\varepsilon_{\text{H}}$) are from a formal
point of view much more appealing than potentials like mBJLDA or HLE16.

\section{\label{sec:overview}Overview of results}

\subsection{\label{sec:bandgap}Fundamental band gap}

The fundamental band gap is very often the quantity one is interested in
when calculating the electronic band structure of a semiconducting or
insulating solid, and $GW$\cite{HedinPR65,AryasetiawanRPP98} is often considered as the
current state-of-the-art method. However, despite recent advances in the
speedup of $GW$ methods,\cite{GovoniJCTC15,GaoSR16,LiuPRB16} the calculations
are not yet routinely applied to large systems. Furthermore, with
one-shot $G_{0}W_{0}$, the results may depend on the
used orbitals, which is particularly the case for antiferromagnetic (AFM)
oxides.\cite{JiangPRB10} Also, the convergence with respect to the number
of unoccupied orbitals may be difficult to achieve.\cite{ShihPRL10,FriedrichPRB11,JiangPRB16}
Alternatively, hybrid functionals can be used, but they are also
very expensive, albeit less than the $GW$ methods.
As already mentioned above, there exist semilocal methods that are able to provide
band gaps with an accuracy that can be comparable to $GW$ or hybrids
depending on the test set.
A certain number of benchmark calculations for large test sets of solids were done,
and below we summarize the results for some of them.

\subsubsection{\label{sec:76solids}Test set of 76 solids}

\begin{table*}
\caption{\label{tab:table_band_gap76}Mean errors with respect to experiment
for the band gap of 76 solids obtained with various DFT methods.
The M(A)E is in eV and the M(A)RE is in \%. The results, obtained with the
\textsc{WIEN2k} code, are from Ref.~\onlinecite{TranPRM18} for BJLDA, LB94,
and GLLB-SC, and Ref.~\onlinecite{TranJPCA17} for all other methods.}
\begin{ruledtabular}
\begin{tabular}{lcccccccccccc} 
           & LDA   & PBE & EV93PW91 & AK13 & Sloc  & HLE16 & BJLDA & mBJLDA & LB94 & GLLB-SC & HSE06 & B3PW91 \\ 
\hline
ME         & -2.17 & -1.99 & -1.55 & -0.28 & -0.76 & -0.82 & -1.53 & -0.30 & -1.87 &  0.20   & -0.68 & -0.36  \\ 
MAE        &  2.17 &  1.99 &  1.55 &  0.75 &  0.90 &  0.90 &  1.53 &  0.47 &  1.88 &  0.64   &  0.82 &  0.73  \\ 
MRE        &   -58 &   -53 &   -35 &    -6 &   -21 &   -20 &   -41 &    -5 &   -54 &    -4   &    -7 &     6  \\ 
MARE       &    58 &    53 &    36 &    24 &    30 &    25 &    41 &    15 &    55 &    24   &    17 &    23  \\ 
\end{tabular}
\end{ruledtabular}
\end{table*}
\begin{figure}
\includegraphics[width=\columnwidth]{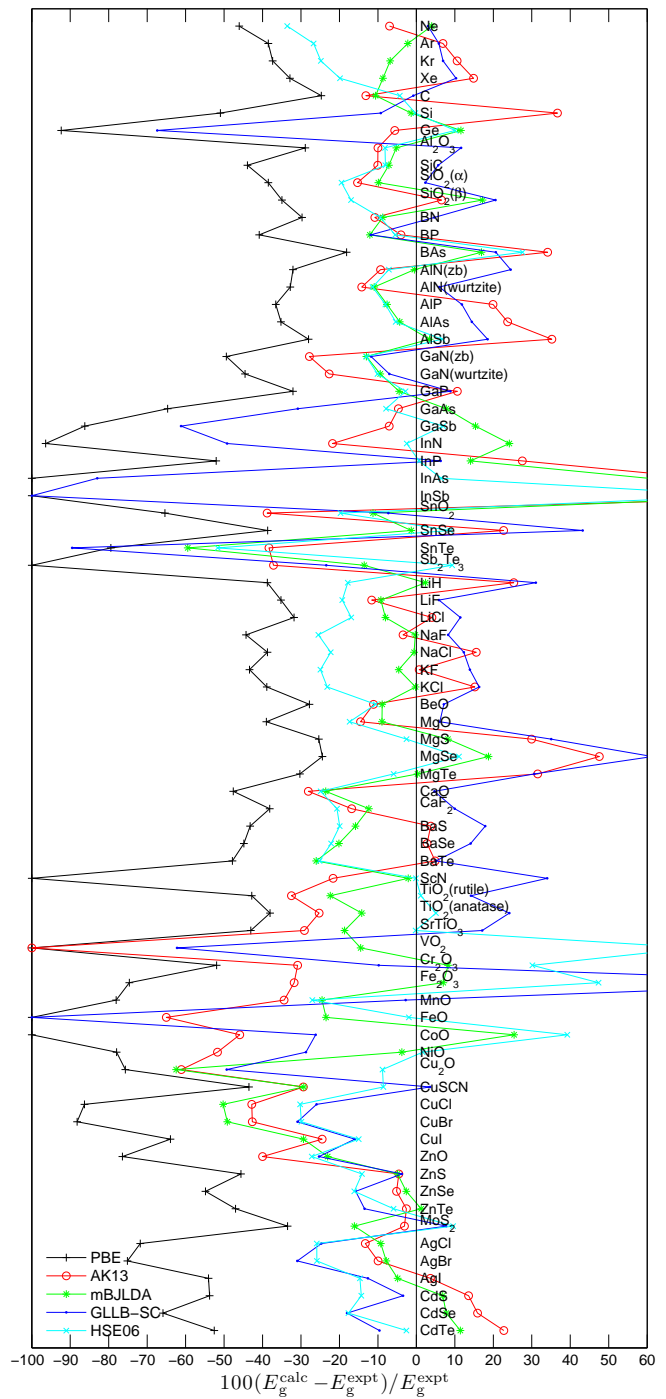}
\caption{\label{fig:fig_band_gap1}Relative error (in \%) with respect to experiment
in the fundamental band gap for the set of 76 solids. The values can be found
in Refs.~\onlinecite{TranJPCA17,TranPRM18}.}
\end{figure}
\begin{figure}
\includegraphics[width=\columnwidth]{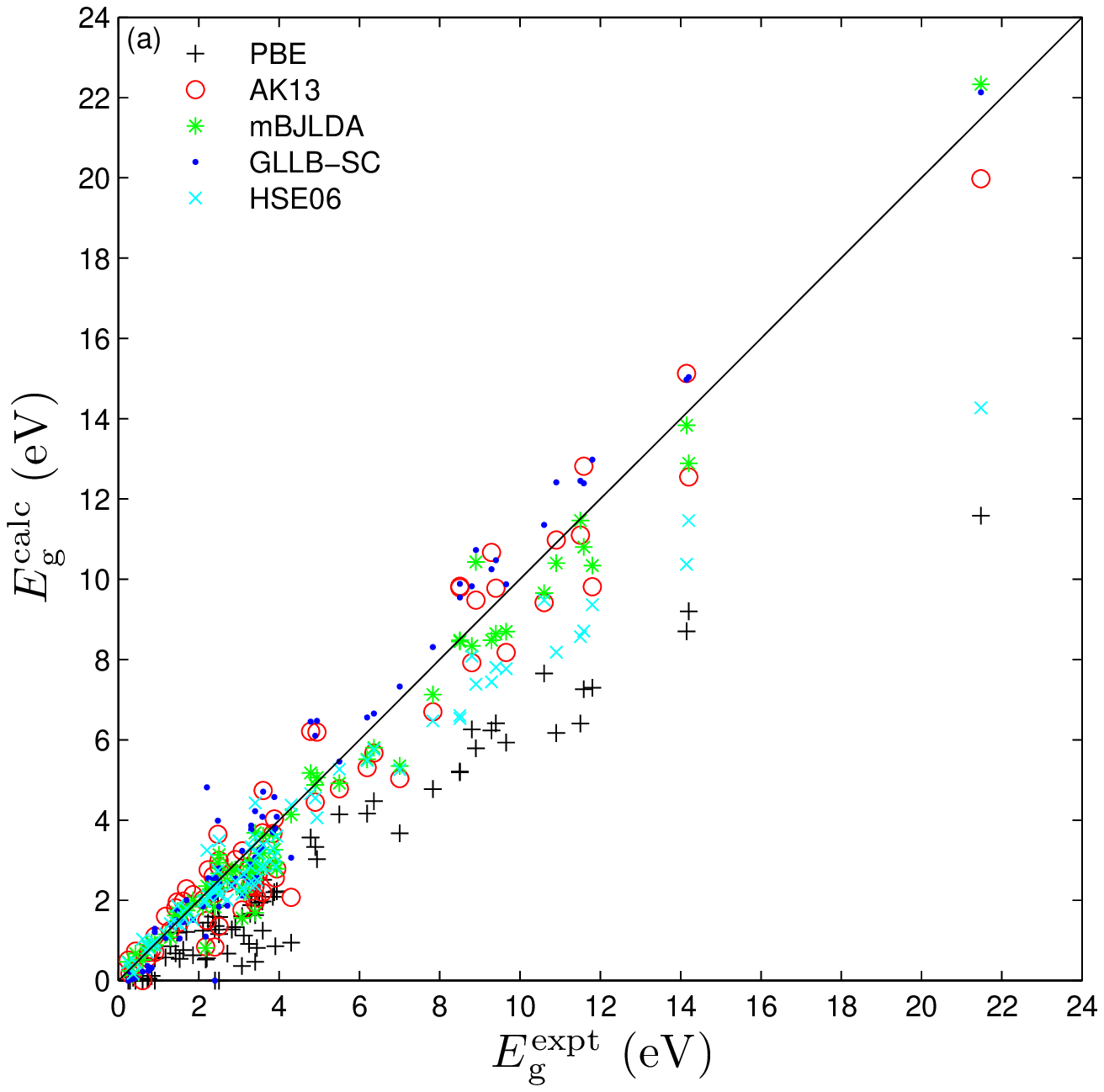}
\includegraphics[width=\columnwidth]{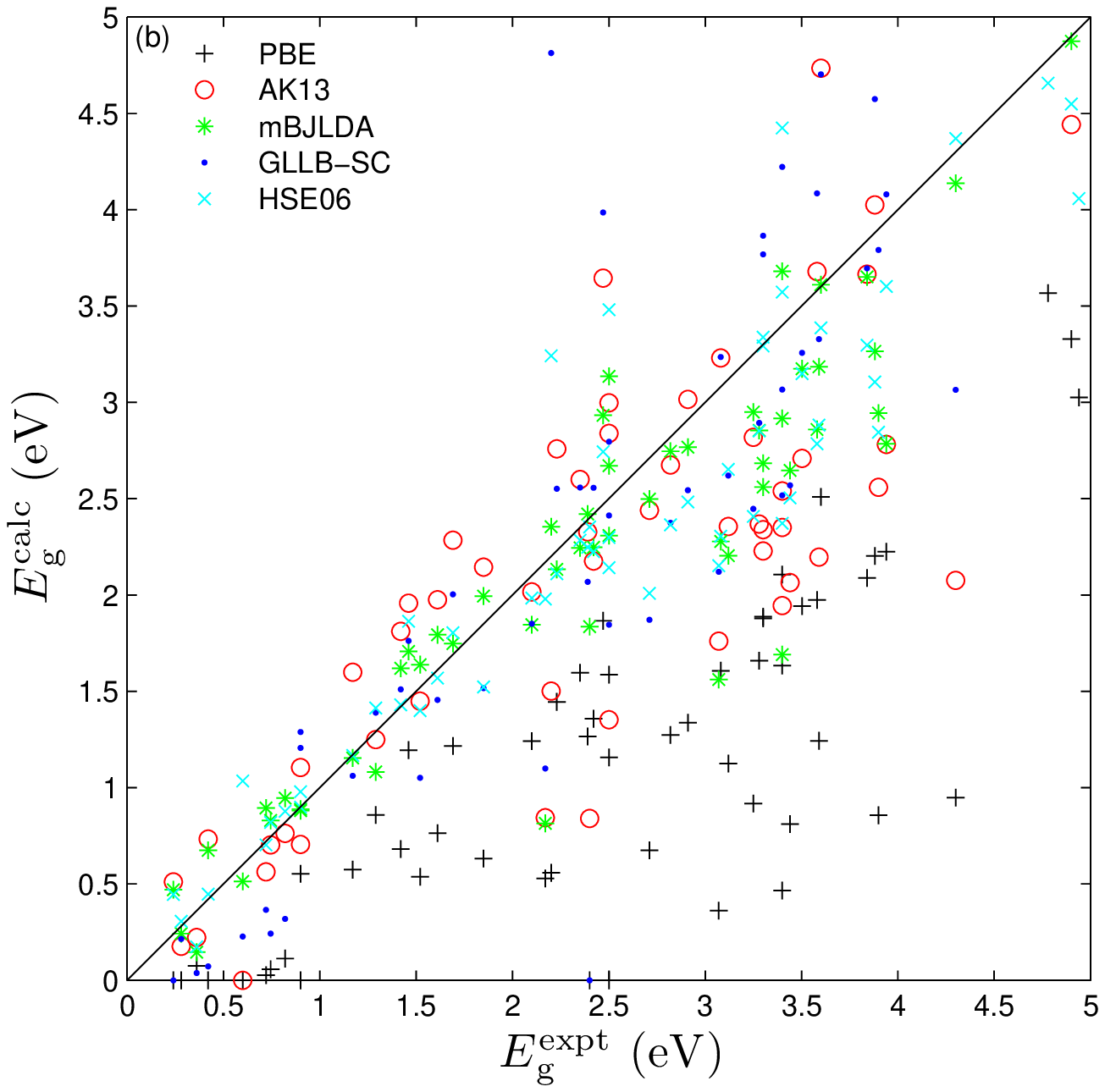}
\caption{\label{fig:fig_band_gap2}Calculated versus experimental fundamental band
gaps for the set of 76 solids. The values can be found in
Refs.~\onlinecite{TranJPCA17,TranPRM18}. The lower panel is a zoom of the upper
panel focusing on band gaps smaller than 5~eV.}
\end{figure}

In two of our previous works,\cite{TranJPCA17,TranPRM18} a set of 76 solids
was build and used for benchmarking and comparing twelve different DFT methods
in total. The tested methods are two LDA-type functionals (LDA and Sloc),
four GGAs (PBE, EV93PW91, AK13, and HLE16), two MGGAs (BJLDA and mBJLDA),
LB94, GLLB-SC, and two hybrids (HSE06 and B3PW91).
The test set (see Fig.~\ref{fig:fig_band_gap1}) consists of
a large variety of semiconductors and insulators, most of them being
IVA solids, IIIA-VA compounds, transition-metal (TM) chalcogenides/halides,
rare gases, or ionic IA-VIIA or IIA-VIA compounds. Among the TM oxides, six of
them (Cr$_{2}$O$_{3}$, Fe$_{2}$O$_{3}$, MnO, FeO, CoO, and NiO)
are AFM with strongly correlated $3d$ electrons.
Also included in the set is VO$_{2}$ that has been extensively studied
(see, e.g, Ref.~\onlinecite{ZhuPRB12b}). Here, the non-magnetic phase of
VO$_{2}$ is considered. The calculations were done with the \textsc{WIEN2k}
all-electron code,\cite{WIEN2k} which is based on the linearized-augmented
plane-wave basis set.\cite{Singh} We mention that the HSE06 results were
actually obtained with YS-PBE0,\cite{TranPRB11} which is also a screened
hybrid functional and was shown to lead to basically the same results as
HSE06 for the electronic structure.\cite{TranPRB11}
As in Refs.~\onlinecite{TranJPCA17,TranPRM18}, we have decided to use
the acronym HSE06.

The mean errors with respect to experiment
\cite{CrowleyJPCL16,LuceroJPCM12,BernstorffOC86,GillenJPCM13,SchimkaJCP11,KollerJPCM13,SkonePRB14,ShiPRB05,LeePRB16,GanoseJMCC16,GrohJPCS09}
are reported in Table~\ref{tab:table_band_gap76}, where M(R)E and MA(R)E denote
the mean (relative) and mean absolute (relative) error, respectively.
Here, we summarize only the
most important observations from Refs.~\onlinecite{TranJPCA17,TranPRM18}.
With a MAE around 2~eV, the most inaccurate xc methods are LDA, PBE, and LB94.
EV93PW91 and BJLDA are only slightly better with $\text{MAE}\sim1.5$~eV.
For the more accurate methods, the ranking depends on whether the MAE or
the MARE is considered, however for both quantities mBJLDA is
the most accurate method with $\text{MAE}=0.47$~eV and $\text{MARE}=15$\%.
The hybrid HSE06 has also a very low MARE (17\%), while
GLLB-SC is the second most accurate method in terms of MAE (0.64~eV).

The detailed results for the 76 solids are shown graphically on
Figs.~\ref{fig:fig_band_gap1} and \ref{fig:fig_band_gap2} for selected methods.
As LDA, PBE systematically underestimates the band gap and this is particularly
severe for small band gaps. mBJLDA does not show a pronounced trend
towards underestimation or overestimation, such that the ME and MRE are among the
smallest. However, mBJLDA does not perform well for the Cu$^{1+}$ compounds
(e.g., Cu$_{2}$O)
and ZnO, but these are basically the only systems for which mBJLDA clearly fails.
While GLLB-SC works much better than mBJLDA for the Cu$^{1+}$ compounds,
it strongly underestimates the band gaps in the heavy IIIA-VA semiconductors
like InSb by nearly 100\%, but also strongly overestimates in a few cases like MgSe and
Fe$_{2}$O$_{3}$. The expensive hybrid HSE06 is very accurate for band gaps
smaller than $\sim7$~eV, but clearly underestimates larger band gaps
[see Fig.~\ref{fig:fig_band_gap2}(a)]. Note that mBJLDA, GLLB-SC, and HSE06
underestimate the band gap of ZnO by a similar amount ($\sim1$~eV).

For the particular case of strongly correlated AFM solids, the mBJLDA results
should be considered as excellent and clearly more accurate than all other
semilocal methods. Only HSE06 is of similar accuracy. Concerning
non-magnetic VO$_{2}$, we mention that in Refs.~\onlinecite{EyertPRL11,ZhuPRB12b}
it is shown that mBJLDA and HSE06 lead to a correct description
of the electronic structure of the (insulating) rutile and
(metallic) monoclinic phases, while LDA and PBE do not.

\subsubsection{\label{sec:472solids}Test set of 472 solids}

\begin{table*}
\caption{\label{tab:table_band_gap472}Mean errors with respect to experiment
for the band gap of 472 solids obtained with various DFT methods.
The M(A)E is in eV and the M(A)RE is in \%. The results are from
Ref.~\onlinecite{BorlidoJCTC19} and were obtained with the
\textsc{VASP} code for all methods, except those for EV93PW91, AK13, and GLLB-SC
that were obtained with the \textsc{WIEN2k} code for the present work.}
\begin{ruledtabular}
\begin{tabular}{lccccccccccc} 
           & LDA  & PBE  & EV93PW91 & AK13 & HLE16 & BJLDA & mBJLDA & SCAN & GLLB-SC & HSE06 & PBE0 \\ 
\hline
ME         & -1.1 & -1.0 & -0.7     & -0.1 & -0.4  & -0.7  & -0.2   & -0.7 &   0.4   & -0.1  &  0.5 \\
MAE        &  1.2 &  1.1 &  0.8     &  0.5 &  0.6  &  0.8  &  0.5   &  0.8 &   0.7   &  0.5  &  0.8 \\
MRE        &  -47 &  -41 &  -22     &    9 &   -7  &  -27  &   -2   &  -27 &    16   &   10  &   53 \\
MARE       &   51 &   46 &   36     &   35 &   32  &   36  &   30   &   38 &    39   &   31  &   61 \\
\end{tabular}
\end{ruledtabular}
\end{table*}

\begin{figure}
\includegraphics[scale=0.58]{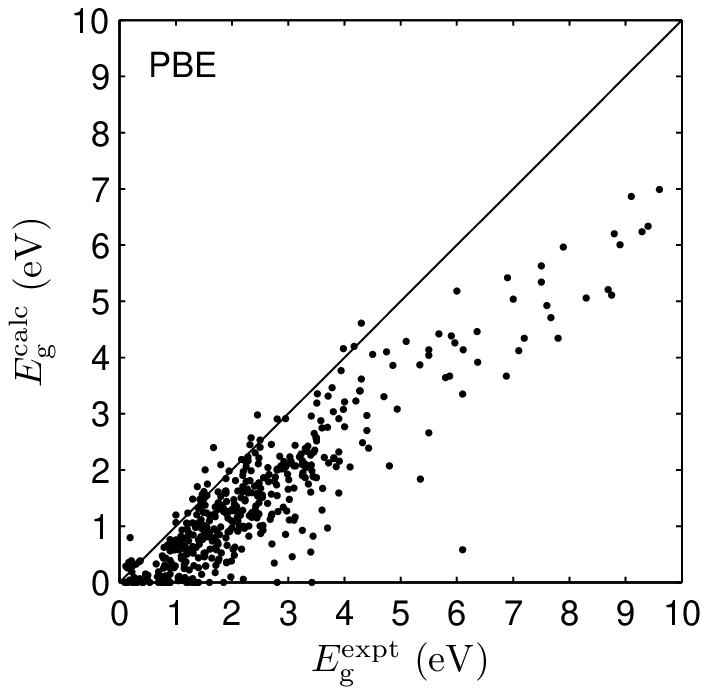}
\includegraphics[scale=0.58]{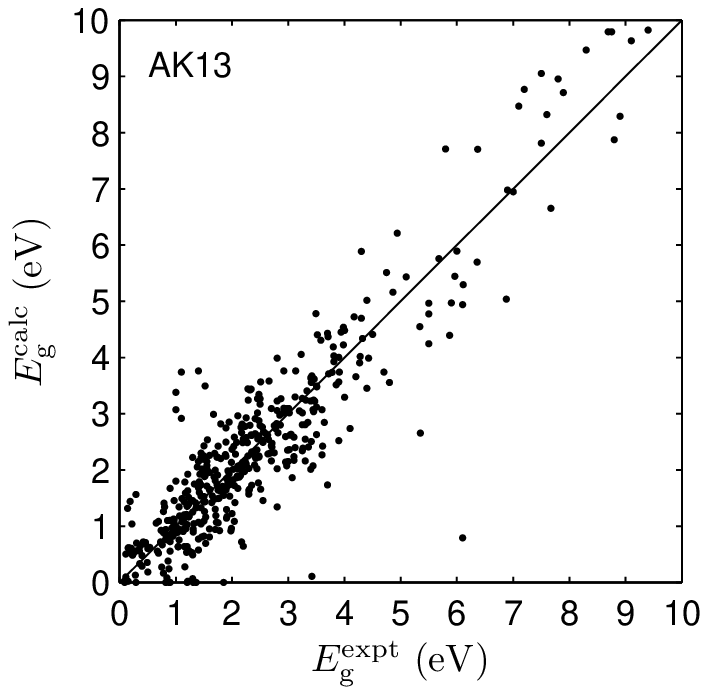}
\includegraphics[scale=0.58]{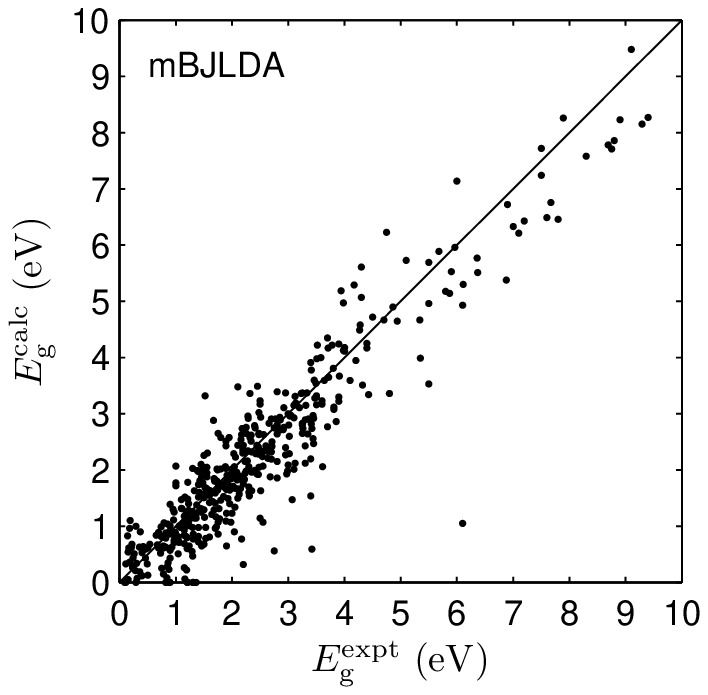}
\includegraphics[scale=0.58]{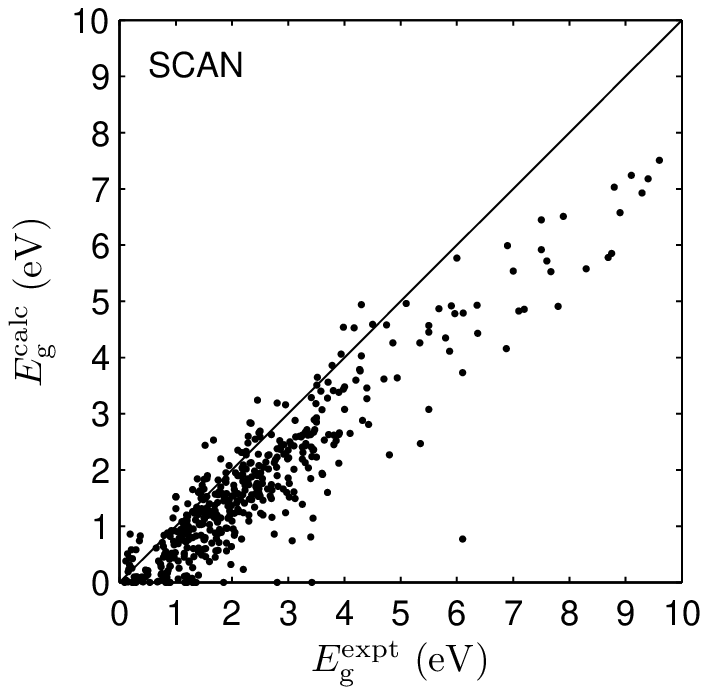}
\includegraphics[scale=0.58]{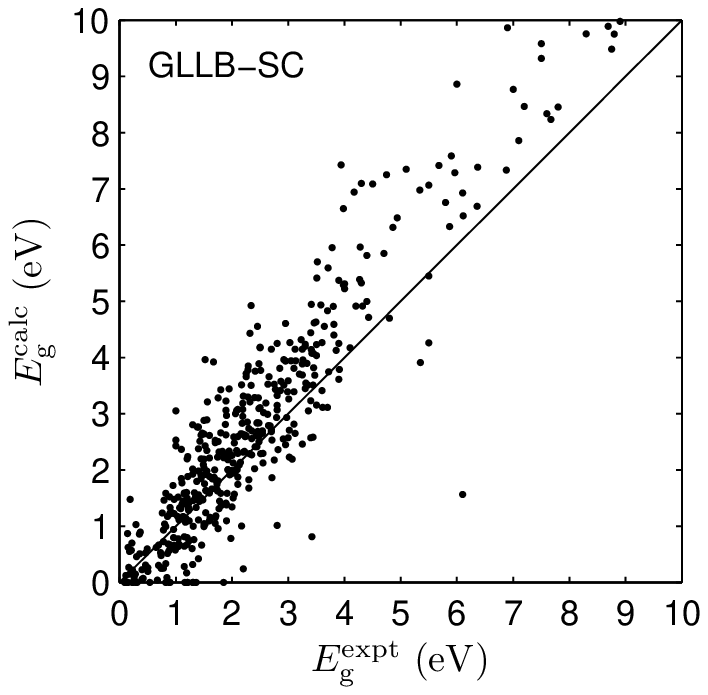}
\includegraphics[scale=0.58]{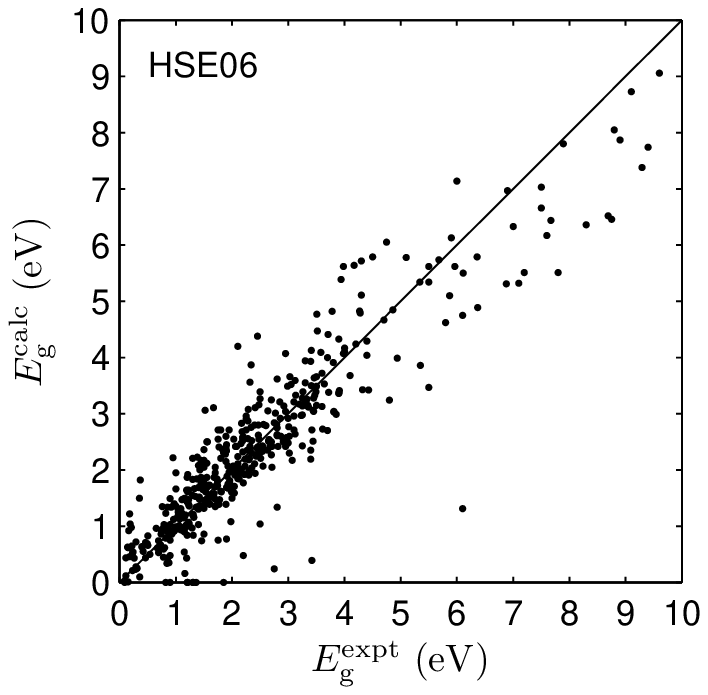}
\caption{\label{fig:fig_band_gap3}Calculated versus experimental fundamental
band gaps smaller than 10~eV for the set of 472 solids.\cite{BorlidoJCTC19}}
\end{figure}

Very recently, a large-scale benchmark study for the band gap was done by some
of us.\cite{BorlidoJCTC19} The data set consists of 472 solids and, as for the
one discussed in the previous section, it consists of a large variety of solids.
Nevertheless, this test set contains no strongly correlated AFM TM oxides.
The calculations from Ref.~\onlinecite{BorlidoJCTC19} were done with the
\textsc{VASP} code, which uses the projector augmented wave
formalism.\cite{BlochlPRB94b,KressePRB96} Additional calculations with the EV93PW91, AK13,
and GLLB-SC (and again PBE) methods were performed for the present work using
the \textsc{WIEN2k} code. The detailed \textsc{WIEN2k} results are in Table~SI
of the supplementary material. Here, we mention that the agreement between
\textsc{VASP} and \textsc{WIEN2k} band gaps is in general excellent for
standard functionals like LDA or PBE. However, for a couple of solids,
those with very large band gaps like the rare gases, noticeably larger
disagreement can be obtained with other xc methods like mBJLDA or HLE16.
In such cases, the discrepancy can be of the order of the eV, which however,
has no real impact on the conclusion.

A summary of the statistics for the error is shown in Table~\ref{tab:table_band_gap472},
and the data for the band gaps smaller than 10~eV are
shown in Fig.~\ref{fig:fig_band_gap3} for selected methods (see Figs.~S1-S11
of the supplementary material for all methods and the full range of band gap values).
The mBJLDA, AK13, HLE16, and HSE06 methods lead to the lowest MAE and MARE,
which are $\sim0.5$~eV and $30$-$35\%$, respectively. These four methods can be
considered as performing equally well. For the test set discussed in
Sec.~\ref{sec:76solids}, the mBJLDA potential was shown to be clearly more
accurate, in particular in terms of MAE.
Two reasons may explain why this is not the case for the large test set.
First, the errors (in eV, but no in \%) with HSE06 are clearly smaller for small
band gaps, and since the test set of 472 solids contains (many) more such small
band gaps (mainly around 2~eV), it favors HSE06. Regarding AK13 and HLE16,
the second reason may be due to the fact that TM oxides, considered in
Sec.~\ref{sec:76solids}, but not here, are very well described by mBJLDA,
but not with AK13 and HLE16.

The other functionals are less accurate. We remark that BJLDA and SCAN
lead to quasi-identical mean errors ($\text{MAE}=0.8$~eV and $\text{MARE}=36$-$38$\%),
and that the screened hybrid HSE06 is much more accurate than the
unscreened PBE0. A much more detailed discussion of the
results can be found in Ref.~\onlinecite{BorlidoJCTC19}.

\subsubsection{\label{sec:bandgap_literature}Other benchmark studies}

Among the xc methods tested for the band gap with the sets of 76 and 472 solids
(discussed above), SCAN is the only one consisting of a MGGA energy functional
with a potential implemented in the gKS framework. Such MGGA methods are
extremely promising for band gap calculation since they are of the
semilocal type (and therefore computationally fast) and lead to a non-zero
xc derivative discontinuity, allowing for a theoretically justified comparison
of the gKS band gap $E_{\text{g}}^{\text{gKS}}$ with the experimental
$E_{\text{g}}$.

As discussed above, SCAN is inferior to the semilocal methods that were designed
specifically for band gaps (mBJLDA and HLE16)
for the set of 472 solids. However, many other MGGA functionals have been proposed,
and several of them have been tested for band gaps in
Refs.~\onlinecite{HeydJCP05,ZhaoJCP09,LuceroJPCM12,PeveratiJCTC12,PeveratiJCP12,PeveratiPCCP12a,XiaoPRB13,YangPRB16,VermaJPCC17,WangPNAS17,MoCPL17,JanaJCP18a,MejiaRodriguezPRB18,PatraPRB19}.
In most of these studies, the test sets are of smaller size (up to 40 solids) and
contain no AFM solids, such that a direct comparison of the mean errors with
our two previous benchmarks is not possible. However, it should be possible
to get a rough idea of the performance of these MGGAs with respect to mBJLDA or
GLLB-SC, which were not considered in these studies.

Truhlar and co-workers have proposed several MGGA functionals that have been
tested on band gaps. Among them, an interesting one is HLE17,\cite{VermaJPCC17}
which shows a performance that is similar to the GGA HLE16 and hybrid
HSE06.\cite{VermaJPCC17} For the three functionals, the MAE is 0.30-0.32~eV for a set
of 31 semiconductors. For this test set, mBJLDA leads to a MAE of 0.27~eV.
However, for the AFM TM monoxides MnO, FeO, CoO, and NiO,
HLE17 is clearly worse than HSE06 and mBJLDA by providing band gaps that are
too small by 1-2~eV with respect to experiment.
Nevertheless, HLE17 is better than SCAN according to the
results in Ref.~\onlinecite{PengPRB17b}.
In Ref.~\onlinecite{WangPNAS17}, the MGGA revM06-L was shown to perform
better than a certain number of other functionals for the same set of 31
semiconductors. However, the MAE is 0.45~eV, which is larger than HLE17.

To date, HLE17 seems to be the most accurate MGGA (implemented within a gKS
framework) according to our knowledge. However, HLE17 is slightly inferior
to mBJLDA for the test set of 31 semiconductors and does not perform well
for the AFM oxides. Other standard MGGA functionals like TPSS\cite{TaoPRL03} or
revTPSS,\cite{PerdewPRL09} or the recent one from Tao and Mo\cite{TaoPRL16} barely
improve over PBE (see, e.g., Ref.~\onlinecite{JanaJCP18a}).
Also noteworthy is a very recent MGGA functional proposed by
Patra \textit{et al}.\cite{PatraPRB19} leading to a MAE of
0.69~eV for a set of 67 solids (comprising no AFM TM oxides), which is,
however, larger than 0.44~eV obtained with mBJLDA for the same test set.

A short summary of other benchmark studies involving the mBJLDA potential
is now given. Jiang\cite{JiangJCP13} compared PBE and mBJLDA on a set
of 50 solids, which includes TM dichalcogenides and Ti-containing oxides.
As expected mBJLDA is much more accurate than PBE. Another goal of this work
was to evaluate the accuracy of the non-self-consistent calculation of the
mBJLDA band gap, which consists of one iteration on top of a PBE calculation
(mBJLDA@PBE). In most cases, the difference between the mBJLDA and
mBJLDA@PBE band gaps is below 0.1~eV. The difference is much larger in rare-gas
solids, e.g., Ne where the band gap with mBJLDA@PBE is smaller than the
self-consistent one by 4.5~eV. Another case is ZnO where this time the mBJLDA@PBE
band gap is larger by $\sim0.5$~eV. It is also found that mBJLDA@PBE is
only slightly less accurate than one-shot $G_{0}W_{0}$ for the TM
dichalcogenides.

Lee \textit{et al}.\cite{LeePRB16} calculated the band gap of 270 compounds
using PBE, mBJLDA, and $G_{0}W_{0}$. They found experimental values for
32 of these compounds, and for this subset the root mean squared error
is 1.77, 0.91, and 0.50~eV for PBE, mBJLDA, and $G_{0}W_{0}$, respectively.

In Ref.~\onlinecite{PandeyJPCC17}, the performance of the GLLB-SC and mBJLDA
potentials were compared on a set consisting of 33 chalcopyrite, kesterite,
and wurtzite polymorphs of II-IV-V$_{2}$ and III-III-V$_{2}$ semiconductors.
The GLLB-SC and mBJLDA band gaps are relatively close in the majority of cases.
Since the experimental values are known for only half of the systems,
a conclusion about the relative accuracy of GLLB-SC and mBJLDA was not really
possible.

In a very recent study, Nakano and Sakai\cite{NakanoJAP18} calculated the
band gap, refractive index, and extinction coefficient of 70 solids with the
PBE and mBJLDA methods. In this work, one of the parametrization of mBJLDA from
Ref.~\onlinecite{KollerPRB12} was used.
Similarly as in the test set of 472 solids discussed in
Sec.~\ref{sec:472solids}, the solids are of various types ranging from simple
systems like diamond or BN to more complicated cases like LiTaO$_{3}$ or
Y$_{3}$Al$_{5}$O$_{12}$. For the band gap, the root mean squared error with
respect to experiment is 0.44 and 1.69~eV, for mBJLDA and PBE, respectively.

In Ref.~\onlinecite{MeinertPRB13}, Meinert reported PBE and mBJLDA calculations on
a set of 26 half-metallic Heusler compounds $X_{2}YZ$, where $X$ and $Y$ are
$3d$ TM, and $Z$ is a main group element. One of his conclusions is that mBJLDA
gives too large band gaps, while PBE only slightly underestimates.
However, as mentioned by Meinert, the experimental measurement of the
half-metallic band gap is difficult and can only be made indirectly.
In addition, such experimental results were available only for a few of the
systems.

On the side of the GLLB-SC potential, we mention Ref.~\onlinecite{CastelliEES12a},
where a MAE of 0.5~eV was obtained for a set of about 30 oxides.
GLLB-SC was then further used for the search of new efficient
photoelectrochemical cells among several thousands of
oxides.\cite{CastelliEES12a,CastelliEES12b,CastelliAEM14}

\subsubsection{\label{sec:further_discussion}Further discussion}

In conclusion, the results published so far have shown that among the
computationally fast (i.e., semilocal) methods, the mBJLDA potential is
overall the most accurate for band gap prediction. Other methods like the
GLLB-SC potential or the MGGA HLE17 may reach similar accuracy for a particular
class of systems. Actually, GLLB-SC performs better than mBJLDA for
Cu$^{1+}$ compounds. However, for the AFM TM solids, mBJLDA is clearly superior
and reaches the accuracy of the much more expensive hybrid functionals.
We mention that the more sophisticated hybrid methods with a fraction
$\alpha_{\text{x}}$ of HF exchange that depends on the dielectric function seem to
be more accurate than traditional hybrids and mBJLDA.\cite{ChenPRM18}

We should also mention that Jishi \textit{et al}.\cite{JishiJPCC14}
pointed out that, although improving over PBE, mBJLDA still
underestimates the band gap in lead halide perovskites by roughly 1~eV
when spin-orbit coupling is included in the calculation.
They proposed a reparametrization [$\alpha=0.4$ and $\beta=1.0$~bohr$^{1/2}$
in Eq.~(\ref{eq:c})] which leads to excellent agreement with experiment,
but would lead to overestimations for other systems (see Sec.~\ref{sec:outlook}).
Similar observations were made in Ref.~\onlinecite{TraorePRB19} about
the more complicated layered hybrid organic-inorganic lead halide perovskites.

\begin{table}
\caption{\label{tab:table_band_gap_f_systems}Fundamental band gap (in eV)
in rare-earth oxides (CeO$_{2}$ and Ce$_{2}$O$_{3}$) and one
actinide oxide (UO$_{2}$). The results were obtained with the WIEN2K code.}
\begin{ruledtabular}
\begin{tabular}{lccc} 
        & CeO$_{2}$ & Ce$_{2}$O$_{3}$ & UO$_{2}$ \\ 
\hline
PBE     & 2.0     & 0.0 & 0.0 \\
HLE16   & 1.4     & 0.2 & 0.0 \\
mBJLDA  & 2.2     & 1.3 & 0.5 \\
GLLB-SC & 3.5     & 1.5 & 0.9 \\
PBE+$U$ & 2.5     & 2.3 & 2.9 \\
Expt.   & 3\footnotemark[1] & 2.4\footnotemark[2] & 2.0\footnotemark[3] \\
\end{tabular}
\end{ruledtabular}
\footnotetext[1]{Ref.~\onlinecite{WuilloudPRL84}.}
\footnotetext[2]{Ref.~\onlinecite{ProkofievJAC96}.}
\footnotetext[3]{Ref.~\onlinecite{IdrissSSR10}.}
\end{table}

In the literature, not much has been said about the accuracy of specialized
potentials like mBJLDA or GLLB-SC for systems with a $4f$ or $5f$ (open) shell
at the VBM or CBM. In order to give a vague idea, we considered CeO$_{2}$ (non-magnetic),
Ce$_{2}$O$_{3}$ (AFM), and UO$_{2}$ (AFM). The results obtained with the
\textsc{WIEN2k} code for a few selected
methods are shown in Table~\ref{tab:table_band_gap_f_systems}, where
in addition to the experimental values, PBE+$U$ results are also given.
For all three systems, the PBE+$U$ calculations were done using the fully
localized limit version\cite{CzyzykPRB94} with $U=5$ and $J=0.5$~eV,
which are similar to suggested values.\cite{DaSilvaPRB07,HeJPCC13}
For CeO$_{2}$, mBJLDA (but not HLE16) slightly improves over PBE,
while GLLB-SC widens further the band gap.
For Ce$_{2}$O$_{3}$ and UO$_{2}$, none of the methods except PBE+$U$ leads
to reasonable band gaps. Furthermore, taking into account spin-orbit coupling
for UO$_{2}$ (not done for the present work) should reduce further the
calculated band gap. Overall, GLLB-SC is more accurate than PBE, HLE16, and mBJLDA,
but less than PBE+$U$, which should be the preferred fast
method for $4f$ and $5f$ systems. Note that the hybrid functional HSE
leads to good agreement with experiment for the three
systems.\cite{HayJCP06,HeJPCC13}

At that point we want to mention other fast DFT schemes for band gap
calculations that we did not consider for the present work.
A few of them do not consist of a particular xc potential,
but of a post-KS-DFT procedure with standard LDA or PBE.
The methods of Chan and
Ceder\cite{ChanPRL10} and Zheng \textit{et al}.\cite{ZhengPRL11} lead to
results similar to mBJLDA for typical $sp$-semiconductors, and actually
more accurate for ZnO, which is a difficult case for mBJLDA.
By making simplifications in the $GW$ equations,
Johnson and Ashcroft\cite{JohnsonPRB98} proposed a shift for the
conduction band. This is somehow in the same spirit as the GLLB-SC method,
but with the disadvantage that the dielectric constant is needed.

\subsection{\label{sec:other_properties}Other properties}

In Sec.~\ref{sec:discontinuity}, it was mentioned that multiplicative
potentials like mBJLDA, AK13, or HLE16 leading to band gaps
$E_{\text{g}}^{\text{KS}}$ close to $E_{\text{g}}$ may possibly lead to
inaccurate results for other properties.
Below, a short summary of the performance of xc potentials on
properties other than the fundamental band gap is given.

\subsubsection{\label{sec:em_bw}Effective mass and bandwidth}

In Refs.~\onlinecite{KimPRB10,TranPRM18} (see also Ref.~\onlinecite{AraujoJAP13}),
the effective hole and electron masses were calculated for five III-V
semiconductors (InP, InAs, InSb, GaAs, and GaSb).
The results showed that HLE16 and mBJLDA increase very often, but not
systematically, the effective masses compared to PBE. With both methods,
the tendency is to yield values that are larger than experiment.
Overall, the results\cite{KimPRB10,TranPRM18} with HLE16 and mBJLDA are
more accurate than with the standard PBE, but less accurate than with the
hybrid HSE (as shown recently in Ref.~\onlinecite{RoedlPRM19} for Ge)
or EV93PW91. GLLB-SC was shown to be not particularly accurate
for the III-V semiconductors.\cite{TranPRM18} The overestimation of the
effective mass by mBJLDA in various perovskites has been been reported in
recent works.\cite{OhkuboJPSJ17,TraorePRB19} In particular, in
Ref.~\onlinecite{TraorePRB19} a reoptimization of the $\alpha$ and $\beta$
parameters in mBJLDA [Eq.~(\ref{eq:c})] specific for lead halide perovskites
and within a pseudopotential implementation has been proposed (see also
Refs.~\onlinecite{JishiJPCC14,JishiAIMSMS16}). It is shown
that the overestimation of the reduced effective mass with the reoptimized
mBJLDA is of similar magnitude as the underestimation with LDA.

As underlined, e.g., in
Refs.~\onlinecite{SinghPRB10b,SmithCMS12,KressePRB12,WaroquiersPRB13,AraujoJAP13,TraorePRB19,WangJPCM19},
a narrowing of electronic bands is observed with mBJLDA compared to LDA/PBE,
which is related to the increase in the effective mass observed above.
On a set of ten cubic semiconductors and insulators,
Waroquiers \textit{et al}.\cite{WaroquiersPRB13} showed that mBJLDA
clearly underestimates the bandwidth with respect to experiment
and is less accurate than LDA. For this test set LDA is, on average,
as accurate as $G_{0}W_{0}$. Actually, the too narrow bands obtained
with mBJLDA have recently been shown to be a source of problem for optical
spectra in ZnSe.\cite{WangJPCM19} A similar reduction in the bandwidth has been
observed with the AK13 potential\cite{VlcekPRB15} and most likely
the same problem occurs with HLE16.

In summary, while mBJLDA is much more accurate than the standard LDA and PBE
for the fundamental band gap, it seems to be of similar accuracy as (or possibly
slightly more accurate than) PBE for the effective mass, but quite inaccurate
for the bandwidth.

\subsubsection{\label{sec:optics}Optics}

The mBJLDA potential has also been used quite frequently for the calculation
of the optical properties of solids. A few representative works reporting
calculations of (non-)linear optics using the RPA for the dielectric
function is given by
Refs.~\onlinecite{SinghPRB10a,KressePRB12,KarsaiPRB14,ReshakJAP16,KopaczekJAP16,OndrackaPRB17,Ibarra-HernandezPRB17,NakanoJAP18,RoedlPRM19}.
Without entering into details, a rather general conclusion
from all these studies is that mBJLDA improves over LDA
and PBE for the optical properties. This is not that surprising since
an improvement in the band gap should in principle be followed
by a more realistic onset in the absorption spectrum.

We also mention a series of studies from Yabana and co-workers
(see, e.g., Refs.~\onlinecite{WachterPRL14,SatoPRB15,SatoJCP15,WachterNJP15,UemotoJCP19,YamadaEPJD19})
who used mBJLDA within time-dependent DFT (TDDFT) to study nonlinear effects
induced by strong short laser pulses.
For instance, in Ref.~\onlinecite{SatoJCP15} they showed that in Si and Ge, mBJLDA and HSE
lead to very similar results (for the dielectric function and excitation energies) and
improve over LDA. However, it was also noticed
that special care is needed in order to get a stable time evolution in TDDFT
when using the mBJLDA potential.\cite{SatoJCP15} Similar problems with the
original BJ potential were reported in Ref.~\onlinecite{KarolewskiPRA13}.

Regarding other xc methods, Vl\v{c}ek \textit{et al}.\cite{VlcekPRB15} reported
an improvement (with respect to PBE) in the dielectric constant when using AK13.
Calculations of the dielectric constant with TDDFT\cite{YanPRB12} showed that
GLLB-SC degrades the results with respect to LDA, in particular if the derivative
discontinuity is taken into account. However, these errors obtained with GLLB-SC
were attributed to the missing electron-hole interaction in the calculation
of the RPA (or adiabatic LDA) response function.

\subsubsection{\label{sec:magnetism}Magnetism}

\begin{table*}
\caption{\label{tab:table_MMI}Atomic spin magnetic moment $\mu_{S}$
(in $\mu_{\text{B}}$) of AFM TM oxides compared to the experimental total
moment $\mu_{S}+\mu_{L}$. The orbital moment $\mu_{L}$ should be in the range
0.6-1~$\mu_{\text{B}}$ for FeO,
\cite{SvanePRL90,TranPRB06,RadwanskiPB08,SchronJPCM13}
1-1.6~$\mu_{\text{B}}$ for CoO,
\cite{SvanePRL90,SolovyevPRL98,ShishidouJPSJ98,NeubeckJPCS01,JauchPRB02,GhiringhelliPRB02,RadwanskiPB04,TranPRB06,RadwanskiPB08,BoussendelPRB10,SchronJPCM13}
0.3-0.45~$\mu_{\text{B}}$ for NiO,
\cite{SvanePRL90,FernandezPRB98,NeubeckJPCS01,RadwanskiPB04,RadwanskiPB08}
and smaller for the other cases. The theoretical results, obtained with the
\textsc{WIEN2k} code, are from Ref.~\onlinecite{TranPRM18}.}
\begin{ruledtabular}
\begin{tabular}{lccccccc}
Method   & MnO  & FeO  & CoO  & NiO  & CuO  & Cr$_{2}$O$_{3}$ & Fe$_{2}$O$_{3}$ \\
\hline
PBE      & 4.17 & 3.39 & 2.43 & 1.38 & 0.38 & 2.44            & 3.53            \\
HLE16    & 4.51 & 3.62 & 2.59 & 1.48 & 0.40 & 2.96            & 4.02            \\
mBJLDA   & 4.41 & 3.58 & 2.71 & 1.75 & 0.74 & 2.60            & 4.09            \\
GLLB-SC  & 4.56 & 3.74 & 2.73 & 1.65 & 0.55 & 2.99            & 4.43            \\
HSE06    & 4.36 & 3.55 & 2.65 & 1.68 & 0.67 & 2.61            & 4.08            \\
Expt.    &
4.58\footnotemark[1] &
3.32,\footnotemark[2]4.2,\footnotemark[3]4.6\footnotemark[4] &
3.35,\footnotemark[5]3.8,\footnotemark[2]\footnotemark[6]3.98\footnotemark[7] &
1.9,\footnotemark[1]\footnotemark[2]2.2\footnotemark[8]\footnotemark[9] &
0.65\footnotemark[10] &
2.44,\footnotemark[11]2.48,\footnotemark[12]2.76\footnotemark[13] &
4.17,\footnotemark[14]4.22\footnotemark[15] \\
\end{tabular}
\end{ruledtabular}
\footnotetext[1]{Ref.~\onlinecite{CheethamPRB83}.} 
\footnotetext[2]{Ref.~\onlinecite{RothPR58}.} 
\footnotetext[3]{Ref.~\onlinecite{BattleJPC79}.} 
\footnotetext[4]{Ref.~\onlinecite{FjellvagJSSC96}.} 
\footnotetext[5]{Ref.~\onlinecite{KhanPRB70}.} 
\footnotetext[6]{Ref.~\onlinecite{HerrmannRonzaudJPC78}.} 
\footnotetext[7]{Ref.~\onlinecite{JauchPRB01}.} 
\footnotetext[8]{Ref.~\onlinecite{FernandezPRB98}.} 
\footnotetext[9]{Ref.~\onlinecite{NeubeckJAP99}.} 
\footnotetext[10]{Ref.~\onlinecite{ForsythJPC88}.} 
\footnotetext[11]{Ref.~\onlinecite{GolosovaJAC17}.} 
\footnotetext[12]{Ref.~\onlinecite{BrownJPCM02}.} 
\footnotetext[13]{Ref.~\onlinecite{CorlissJAP65}.} 
\footnotetext[14]{Ref.~\onlinecite{BaronSSS05}.} 
\footnotetext[15]{Ref.~\onlinecite{HillCM08}.} 
\end{table*}

Concerning magnetism, it has been shown that the mBJLDA potential provides
very accurate values of the atomic magnetic moment in AFM systems with
localized electrons. In Refs.~\onlinecite{TranPRL09,KollerPRB11,BotanaPRB12,TranPRM18},
we showed that mBJLDA increases by 0.2-0.4~$\mu_{\text{B}}$ the atomic spin
moment $\mu_{S}$ with respect to PBE. Since PBE quasi-systematically underestimates
$\mu_{S}$ in AFM systems, such an increase leads to better agreement with experiment.
In Table~\ref{tab:table_MMI}, results from Ref.~\onlinecite{TranPRM18} for
AFM TM oxides are reproduced for selected xc methods. Compared to PBE, all other
methods lead to larger $\mu_{S}$ values. Since there is sometimes a rather large
uncertainty in the experimental value and, furthermore, the orbital component
$\mu_{L}$ is not known precisely, it is difficult to say which method is the best.
Anyway, in most cases the agreement with experiment is much improved compared to
PBE and satisfactory. We just note that HLE16 seems to underestimate (overestimate)
the value in CuO (Cr$_{2}$O$_{3}$) and that GLLB-SC overestimates for
Cr$_{2}$O$_{3}$ and Fe$_{2}$O$_{3}$. The advantage of these methods over
DFT+$U$,\cite{AnisimovPRB91} which is widely used for AFM oxides,
is that they do not contain a system-dependent parameter like $U$.

\begin{table}
\caption{\label{tab:table_MMTOT}Unit cell spin magnetic moment $\mu_{S}$
(in $\mu_{\text{B}}$/atom) of $3d$ TM. The experimental values are also spin
magnetic moments. The theoretical results, obtained with the \textsc{WIEN2k}
code, are from Ref.~\onlinecite{TranPRM18}.}
\begin{ruledtabular}
\begin{tabular}{lccc}
Method   & Fe   & Co   & Ni   \\
\hline
PBE      & 2.22 & 1.62 & 0.64 \\
HLE16    & 2.72 & 1.72 & 0.63 \\
mBJLDA   & 2.51 & 1.69 & 0.73 \\
GLLB-SC  & 3.08 & 1.98 & 0.81 \\
HSE06    & 2.79 & 1.90 & 0.88 \\
Expt.    &
1.98,\footnotemark[1]2.05,\footnotemark[2]2.08\footnotemark[3] &
1.52,\footnotemark[3]1.58,\footnotemark[2]\footnotemark[4]1.55-1.62\footnotemark[1] &
0.52,\footnotemark[3]0.55\footnotemark[2]\footnotemark[5] \\
\end{tabular}
\end{ruledtabular}
\footnotetext[1]{Ref.~\onlinecite{ChenPRL95}.}
\footnotetext[2]{Ref.~\onlinecite{Scherz03}.}
\footnotetext[3]{Ref.~\onlinecite{ReckPR69}.}
\footnotetext[4]{Ref.~\onlinecite{MoonPR64}.}
\footnotetext[5]{Ref.~\onlinecite{MookJAP66}.}
\end{table}

Table~\ref{tab:table_MMTOT} shows the results for $\mu_{S}$ in the
ferromagnetic metals Fe, Co, and Ni (results from Ref.~\onlinecite{TranPRM18}).
Unlike in AFM systems with localized electrons, the LDA and standard GGAs lead
usually to reasonable values of the moment in itinerant metals.
Indeed, PBE only slightly overestimates with respect to experiment, while
the other methods clearly overestimate the values.
For the three metals, GLLB-SC and HSE06 overestimate more than
mBJLDA and HLE16. We mention that the reverse has been observed by
Singh\cite{SinghPRB10b} in the ferromagnetic metal Gd, a $4f$-system,
in which the mBJLDA moment is smaller than the PBE one and than the experimental
value by $\sim1$~$\mu_{\text{B}}$. Thus, contrary to what has been observed
in $3d$ magnetic systems, the mBJLDA potential reduces the exchange
splitting in Gd (due to an increase of occupancy of the minority $4f$ states)
and shifts the $spd$-bands up with respect to the $4f$ bands.

In the work of Meinert\cite{MeinertPRB13} on half-metallic Heusler compounds
already mentioned in Sec.~\ref{sec:bandgap_literature}, it is shown that the
PBE and mBJLDA magnetic moments are the same in most cases, which is simply
due to the half-metallic state. Only for Co$_{2}$FeSi and Co$_{2}$FeGe the
moment is clearly different with mBJLDA (larger by $\sim0.35$~$\mu_{\text{B}}$)
and actually in better agreement with experiment.

Regarding the performance of MGGA functionals implemented with a non-multiplicative
potential, a certain number of results are available for ferromagnetic metals.
\cite{SunPRB11b,IsaacsPRM18,JanaJCP18a,RomeroEPJB18,EkholmPRB18,FuPRL18,FuPRB19,MejiaRodriguezPRB19}
From these studies the most interesting results concern the SCAN functional,
which has been shown to overestimate the magnetic moment, and sometimes by a
rather large amount (e.g., $\sim0.5$~$\mu_{\text{B}}$ for Fe). However, such
overestimation is not systematically observed with functionals of this class,
since other MGGA functionals like TPSS\cite{TaoPRL03} or revTPSS\cite{PerdewPRL09}
lead to values similar to PBE.\cite{SunPRB11b,JanaJCP18a,FuPRB19}

In summary, the standard LDA and GGA lead to magnetic moments which
are qualitatively correct for itinerant metals, but not for AFM solids
with strongly correlated electrons (strong underestimation).
On the other hand, methods giving much more reasonable magnetic moments in AFM
solids, e.g., mBJLDA, GLLB-SC, or HSE06, lead to strong overestimation in
itinerant metals. However, it is important to note that the overestimation
of the moment in Fe, Co, and Ni with mBJLDA is much less severe than with
GLLB-SC and HSE06 (an explanation is provided in Sec.~\ref{sec:outlook}).
Note that the inadequacy of hybrid functionals for metals has been
documented.\cite{PaierJCP06,TranPRB12,JangJPSJ12,GaoSSC16}
Currently, there is no DFT method that is able to provide qualitatively
correct values of the moment in itinerant metals and AFM solids
simultaneously.

\subsubsection{\label{sec:electron_density}Electron density}

The quality of the electron density $\rho$ can be measured by considering
the electric field gradient (EFG) or the X-ray structure factors.
The EFG is defined as the second derivative of the Coulomb potential
at a nucleus and, therefore, depends on the charge distribution in the solid.
In Ref.~\onlinecite{TranPRM18}, the calculated EFG of seven metals
(Ti, Zn, Zr, Tc, Ru, Cd, and Cu$_{2}$Mg) and two non-metals
(CuO and Cu$_{2}$O) has been calculated and compared to
experiment. For some of the systems, the error bar on the experimental value
is quite large, nevertheless it was possible to draw some conclusions
on the accuracy of the methods. The method that is the most accurate
overall is the GLLB-SC potential that leads to reasonable errors for
all systems. The methods that were shown to be quite inaccurate
are LDA, Sloc, HLE16, mBJLDA, and LB94.

In the same work,\cite{TranPRM18} the calculated and experimental
X-ray structure factors of Si were compared. The best method
is the screened hybrid HSE06, followed by PBE, EV93PW91, and
BJLDA. The mBJLDA potential is less accurate than these methods,
but slightly more accurate than LDA. The Sloc and HLE16 were shown
to be much more inaccurate than all other methods, while
LB94 and GLLB-SC are also very inaccurate, but to a lesser extent.
Actually, in the case of GLLB-SC it is the core density that is
particularly badly described, while the valence density
seems to be rather accurate (see Ref.~\onlinecite{TranPRM18} for details),
which is consistent with the very good description of the EFG mentioned above.

In an interesting recent study,\cite{MedvedevS17} it was shown that functionals
like SCAN or HSE06 which were constructed more from first-principles by using mathematical
constraints rather than empirically by fitting a large number of coefficients
produce more accurate electron density, however the test set consists only
of atoms/cations with 2, 4, or 10 electrons.
Highly parametrized functionals like those
of the Minnesota family\cite{ZhaoTCA08} are much less accurate.

To finish this section, we mention a study on VO$_{2}$, where various 
functionals are compared for the electron density. Accurate electron
densities from Monte-Carlo calculations were used as reference.
It is shown that hybrid functionals like HSE are more accurate than
LDA and PBE.

\section{\label{sec:summary}Summary}

From the overview of the literature results presented in Sec.~\ref{sec:overview},
the main conclusions are the following:
\begin{itemize}
\item Among the semilocal methods, the mBJLDA potential is on average the most
accurate for the fundamental band gap.
Furthermore, mBJLDA is at least as accurate as hybrid functionals like HSE06.
Only the most advanced methods, namely the dielectric-function-dependent hybrid
functionals and $GW$, can be consistently more accurate than mBJLDA.
\item A particularly strong advantage of mBJLDA over the other methods is its
reliability for all kinds of systems. mBJLDA is more or less equally accurate
for large band gap insulators (ionic solids, rare gases), $sp$-semiconductors,
and systems with TM atoms including AFM systems with localized $3d$-electrons.
\item All other semilocal potentials are very unreliable for the band gap
in AFM systems.
\item The cases where mBJLDA is clearly not accurate enough are the
Cu$^{1+}$ compounds (for which GLLB-SC works much better),
but also ZnO, where the band gap is still underestimated by at least 1~eV.
Note that a modification of mBJLDA, called the universal correction,\cite{RasanenJCP10}
may help for Cu$^{1+}$ compounds (see Ref.~\onlinecite{TranPRB15} for details).
Lead halide perovskites require
another set of parameters $\alpha$ and $\beta$ in mBJLDA,
while none of the multiplicative potentials leads to meaningful
band gaps for the $4f$ and $5f$ systems.
\item The bandwidth is not reproduced accurately by mBJLDA which makes the
bands too narrow. LDA and PBE are more accurate than mBJLDA.
\item The magnetism in AFM systems is very well described with mBJLDA and as accurately
as with hybrid functionals. In ferromagnetic metals the magnetic moment
is too large, however the overestimation is not as large as with GLLB-SC and
hybrid functionals.
\item The electron density does not seem to be particularly well described
by mBJLDA. The standard PBE is more accurate, while HLE16 and GLLB-SC are
extremely inaccurate. HSE06 is particularly good for the electron density of Si.
\item Beside mBJLDA, DFT+$U$ is the only other computationally cheap method
that is able to provide qualitatively correct results in AFM systems for the
band gap and magnetic moment.
\end{itemize}
Thus, this summary shows that the mBJLDA potential is currently the best
alternative to the much more expensive hybrid or $GW$ methods.
This explains why it has been implemented in various codes
\cite{KimPRB10,CesarAPL11,MarquesCPC12,GermaneauCPC13,MeinertPRB13,YePRB15,TraorePRB19,BartokPRB19,Smidstrup19}
and used for numerous applications.
Examples are the search for efficient thermoelectric materials,
\cite{ParkerPRB12,ParkerPRL13,PardoPRB13,ShiPRA15,ZhangNC16,XingPRM17,KangPRM18,LiuPRB18,HeJACS18,FuPRM18,ZhuNC19a,ZhuNC19b,MatsumotoJAP19,TerashimaJAP19}
topological insulators,
\cite{FengPRB10,FengPRL11,FengPRB12,ZhuPRB12a,AgapitoPRL13,ShengPRB14,KufnerPRB15,ShiPRA15,LiSR15,ArnoldNC16,KangPRL16,BrydonPRL16,RuanPRL16,CaoPRB17,RongSR17,ArnoldPRL17,QuanPRL17,OinumaPRB17,ZhangJPCL17,KimSA18,ParveenPCCP18,TangN19}
or materials for photovoltaics.
\cite{LiPRM17,ZhangPRA18,LiJMCC19,OngSR19}
The mBJLDA potential has also been used for the screening of materials
or in relation to machine learning techniques
\cite{LeePRB16,ChoudharySD18,ChoudharyCM19} (see
Refs.~\onlinecite{CastelliEES12a,CastelliEES12b,CastelliAEM14,MiroCSR14,RasmussenJPCC15,ParkPCCP19}
for such works using the GLLB-SC potential).
Furthermore, it has been shown to be accurate for the calculation of
the ELNES (energy loss near-edge structure) spectrum of NiO,\cite{HetabaPRB12} and used
recently for the band gap calculation of configurationally
disordered semiconductors.\cite{XuJCP19}
Nevertheless, as a technical detail we should mention that
the number of iterations to achieve self-consistent field convergence
is usually (much) larger with the mBJLDA potential than with other potentials.

\section{\label{sec:outlook}Outlook}

The question is why mBJLDA is more accurate than all other semilocal methods.
The main reason is the use of two ingredients: the \textit{kinetic-energy
density $t_{\sigma}$} and the \textit{average of $\nabla\rho/\rho$ in the unit
cell $g$} [Eq.~(\ref{eq:g})], that are now discussed.

As illustrated and discussed in previous works,\cite{TranJPCM07,TranJPCA17,TranPRM18}
the mBJLDA potential is, compared to the PBE potential, more negative around
the nuclei (where the orbital of the VBM is usually located) and less
negative in the interstitial region (where the orbital of the CBM is located).
This enlarges the band gap. This mechanism, which works in most ionic solids and
typical $sp$-semiconductors, is also used by the GGAs AK13, HLE16, and Sloc.
However, these GGAs are not accurate for AFM systems, where the band gap may
have a strong onsite $d$-$d$ character. In Ref.~\onlinecite{KollerPRB11}
it is shown that the second term in Eq.~(\ref{eq:mBJ}), which is proportional to
$\sqrt{t_{\sigma}/\rho_{\sigma}}$, allows the potential to act
differently on orbitals with different angular distributions
(e.g., $t_{2g}$ versus $e_{g}$ of the $3d$ states) and increase a $d$-$d$
band gap. In order to do this, such a term in the mBJLDA potential is much
more efficient than what GGA potentials can do.
Actually, the $t_{\sigma}$-dependency in mBJLDA is strong, and most likely
much stronger than in any other MGGA like SCAN or HLE17, which are not as good
as mBJLDA for AFM solids.

\begin{figure}
\includegraphics[width=\columnwidth]{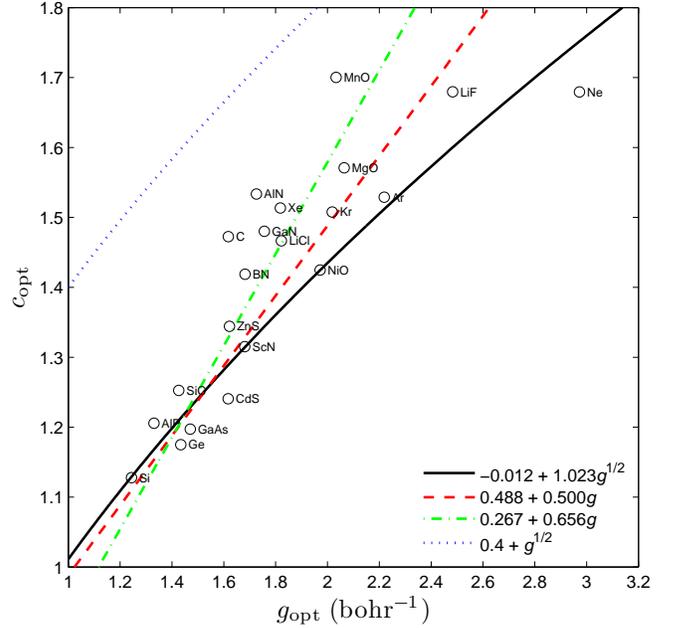}
\caption{\label{fig:fig_mBJ_gropt_copt}Plot of $c_{\text{opt}}$ versus
$g_{\text{opt}}$ (see text for explanations).
The fit of these data points\cite{TranPRL09} given by Eq.~(\ref{eq:c}) is
represented by the solid curve. The two fits (of different sets of solids)
proposed in Ref.~\onlinecite{KollerPRB12} are represented by the dashed and
dashed-dotted curves, and the fit from Ref.~\onlinecite{JishiJPCC14}
specific for lead halide perovskites is represented by the dotted curve.}
\end{figure}
\begin{table*}
\caption{\label{tab:table_g}Results obtained from mBJLDA calculations.
Value of $g=g_{\text{SC}}$ [Eq.~(\ref{eq:g})] (in bohr$^{-1}$)
obtained at the end of a usual self-consistent (SC) calculation.
Fundamental band gap (in eV) and spin magnetic moment $\mu_{S}$ (in $\mu_{\text{B}}$)
obtained from a usual calculation ($g=g_{\text{SC}}$) or with $g$ fixed
($g=g_{\text{fixed}}$) to the value obtained for the corresponding AFM oxide
(for a metal) or metal (for a AFM oxide). The spin magnetic moment is the value in
the unit cell and per atom for metals and inside the TM atomic sphere for the
AFM oxides.}
\begin{ruledtabular}
\begin{tabular}{lcccccccc}
 & Fe & Co & Ni & Cu & FeO & CoO & NiO & CuO \\
\hline
\multicolumn{9}{c}{$g_{\text{SC}}$} \\
 & 1.40 & 1.48 & 1.54 & 1.52 & 1.92 & 1.95 & 1.97 & 1.93 \\
\multicolumn{9}{c}{Fundamental band gap} \\
With $g=g_{\text{SC}}$    & 0    & 0    & 0    & 0 & 1.84 & 3.14 & 4.14 & 2.27 \\
With $g=g_{\text{fixed}}$ & 0    & 0    & 0    & 0 & 1.06 & 1.99 & 3.09 & 1.54 \\
\multicolumn{9}{c}{$\mu_{S}$} \\
With $g=g_{\text{SC}}$    & 2.51 & 3.38 & 0.73 & 0 & 3.58 & 2.71 & 1.75 & 0.74 \\
With $g=g_{\text{fixed}}$ & 2.53 & 3.43 & 0.80 & 0 & 3.49 & 2.63 & 1.67 & 0.67 \\
\end{tabular}
\end{ruledtabular}
\end{table*}

The integral given by Eq.~(\ref{eq:g}), $g$, is even more crucial in explaining
the success of the mBJLDA potential. Figure~\ref{fig:fig_mBJ_gropt_copt} shows
a plot of $c_{\text{opt}}$ versus $g_{\text{opt}}$, where $c_{\text{opt}}$
is the value of $c$ in Eq.~(\ref{eq:c}) that would exactly lead to the experimental
band gap and $g_{\text{opt}}$ is the value of $g$ obtained at the end of the
self-consistent mBJLDA calculation done with $c_{\text{opt}}$.
This is shown for all the solids, except two (FeO and ZnO), that were
considered in our original work.\cite{TranPRL09} A correlation between
$c_{\text{opt}}$ and $g_{\text{opt}}$ is clearly visible and the fit of these
data that we chose in Ref.~\onlinecite{TranPRL09} is given by Eq.~(\ref{eq:c})
with $\alpha=-0.012$ and $\beta=1.023$~bohr$^{1/2}$. Alternative
fits\cite{KollerPRB12} that were determined by using different sets of solids,
e.g., focusing on small band gaps are also shown. This rather clear correlation
between $c_{\text{opt}}$ and $g_{\text{opt}}$ allows for a meaningful fit $c(g)$,
which, when used in the mBJLDA potential, produces accurate band gaps.
However, as mentioned in Sec.~\ref{sec:overview}, the standard parametrization
of $c$ is not good enough for lead halide perovskites, which require
other values of $\alpha$ and $\beta$, and as shown in
Fig.~\ref{fig:fig_mBJ_gropt_copt} this parametrization would lead to
an overestimation of the band gap for other solids (in general, the band gap
increases in a monotonous way with respect to $c$).
Another illustration of the importance of $g$ is given in Table~\ref{tab:table_g},
where the value of $g$ for itinerant $3d$ TM and their oxides is shown.
There is a clear difference between the metals and the oxides.
While the value of $g$ is in the range 1.40$-$1.54~bohr$^{-1}$ for the metals,
it is much larger, 1.92$-$1.97~bohr$^{-1}$, for the monoxides.
From a more detailed analysis, we could see that this difference
comes more particularly from the interstitial region, which is on average
more inhomogeneous (larger $\left\vert\nabla\rho\right\vert/\rho$)
for the oxides. However, also around the TM atom
$\left\vert\nabla\rho\right\vert/\rho$ is larger for the oxides, which
therefore also contributes in making $g$ larger compared to the metal.
If the calculation for the TM oxides is done by fixing the value of
$g$ in Eq.~(\ref{eq:c}) to the value of the corresponding metal, then
a large decrease of 0.7$-$1.1~eV in the band gap (and a large discrepancy with
experiment) is obtained (Table~\ref{tab:table_g}). The same is observed for the
magnetic moment, which is reduced in the oxides when the $g$ of the metal is used.
Conversely, using the value of $g$ from the oxide for the calculation on the
metals leads to an increase of the magnetic moment. The important point to note
is that, in any case, using the incorrect $g$ (i.e., the one from the other
system) leads to much larger errors. We mention that other works using an average
in the unit cell involving the density can be found in
Refs.~\onlinecite{ClarkPRB10,MarquesPRB11,BorlidoJCTC18,TerentjevPRB18a,WangJCP19}.

However, since the mBJLDA potential is certainly not providing an optimal band
gap in every case, and the results for other properties like the bandwidth or
the electron density are even worse than with the standard PBE, there is room
for improvement. As already mentioned in this work, forcing a multiplicative
potential to give KS band gaps $E_{\text{g}}^{\text{KS}}$ close to the experimental
$E_{\text{g}}$ is not really correct from the formal point of view and
may result in a potential that is inappropriate for other properties.
A derivative discontinuity $\Delta_{\text{xc}}$ should be present somewhere in
the theory. $\Delta_{\text{xc}}$ is included in the gKS band gap
$E_{\text{g}}^{\text{gKS}}$ with non-multiplicative potentials (MGGA and
hybrid) or obtained from a post-KS calculation with the GLLB-SC potential.

Thus, from the discussion above, the strategy to follow for
the construction of a \textit{computationally fast} and
\textit{generally accurate} xc potential that alleviates the problems
encountered with mBJLDA is quite obvious:
\begin{itemize}
\item The potential should be based on one of these two types:
non-multiplicative MGGA or multiplicative of the GLLB-SC type.
Then, a derivative discontinuity is available as it should be.
\item Since both $t_{\sigma}$ and $g$ were shown to be crucial for the
mBJLDA potential, most likely they can be useful for other types of
potentials (by definition $t_{\sigma}$ is used in a MGGA). Note that
instead of $g$, a variant may be used, i.e., the average in the unit cell
of a quantity different from $\nabla\rho/\rho$ may be more useful.
\end{itemize}

However, we mention that attempts (presented in Ref.~\onlinecite{TranPRM18}
or unpublished) to improve over the GLLB-SC potential along these lines have been
unsuccessful up to now.
For instance, one of the most sophisticated variants of
Eq.~(\ref{eq:vxcGLLBSC}) that we have considered is of the form (the formula
for $\Delta_{\text{x}}$ should be modified accordingly)
\begin{eqnarray}
v_{\text{xc},\sigma}(\mathbf{r}) & = &
cv_{\text{x},\sigma}^{\text{BR}}(\mathbf{r}) +
\left(3c-2\right)F_{\sigma}(\mathbf{r})K_{\text{x}}^{\text{LDA}} \nonumber \\
& & \times\sum_{n,\mathbf{k}}w_{n\mathbf{k}\sigma}
\sqrt{\epsilon_{\text{H}}-\epsilon_{n\mathbf{k}\sigma}}
\frac{\left\vert\psi_{n\mathbf{k}\sigma}(\mathbf{r})\right\vert^{2}}{\rho_{\sigma}(\mathbf{r})}
+ v_{\text{c},\sigma}^{\text{PBEsol}}(\mathbf{r}), \nonumber \\
\label{eq:vxcGLLBvar}
\end{eqnarray}
where $F_{\sigma}$ is a function that depends and $\rho_{\sigma}$,
$\nabla\rho_{\sigma}$, and $t_{\sigma}$ and should satisfy $F_{\sigma}=1$
for a constant electron density in order to recover the homogeneous electron gas limit,
and $c$ is a function that depends on some average in the unit cell similar
to Eq.~(\ref{eq:c}). With Eq.~(\ref{eq:vxcGLLBvar}), that is of course inspired by
the mBJLDA potential [Eq.~(\ref{eq:mBJ})], we have not been able to find a
strategy to parametrize $c$ such that a clear correlation like in
Fig.~\ref{fig:fig_mBJ_gropt_copt} for mBJLDA is obtained.

There are also a few drawbacks or challenges that should be mentioned:
\begin{itemize}
\item Modelizing a potential directly without requiring that it is a functional
derivative is much easier and leads to much more flexibility. However, such stray
potentials lead to problems\cite{GaidukJCP09,GaidukJCTC09,KarolewskiPRA13} and
do not allow total-energy calculations. Thus, having a potential that is
the derivative of a functional $E_{\text{xc}}$ would be certainly much more
interesting, but also much more challenging. AK13 and HLE16 satisfy this
requirement, but the energy $E_{\text{xc}}$ is very
inaccurate.\cite{CerqueiraJCTC14,LindmaaPRB16,VermaJPCL17}
\item The use of the average in the unit cell of some quantity, e.g. $g$
[Eq.~(\ref{eq:g})], works only in the case of bulk solids. Calculating such
averages for systems with vacuum (surface or molecule)
or interfaces makes no sense. Possible solutions to this problem have been proposed
in Refs.~\onlinecite{MarquesPRB11,BorlidoJCTC18} and consist of taking
the average not over the unit cell, but over a region localized around
the position $\mathbf{r}$ where the potential is calculated.
However, it is not clear how far such region should extend to bring
enough nonlocal information.
\item The use of the highest occupied and lowest unoccupied orbital energies
like in the GLLB-SC potential [Eqs.~(\ref{eq:vxcGLLBSC}) and (\ref{eq:deltax})]
is also problematic. For instance, the values of
$\epsilon_{\text{H}}$ and $\epsilon_{\text{L}}$ for an interface correspond
to one of the two bulk solids, while they correspond to the respective
bulk solids when they are treated separately. This is somehow inconsistent.
The solution would be to define position-dependent functions replacing
$\epsilon_{\text{H}}$ and $\epsilon_{\text{L}}$.
\end{itemize}

Trying to overcome these problems makes the search of a generally applicable
xc potential even more difficult, however, as a first step they can be ignored.

\section*{Supplementary Material}

See supplementary material for the detailed results for the band gap
of the 472 solids discussed in Sec.~\ref{sec:472solids}.

\begin{acknowledgments}

F.T., J.D., and P.B. acknowledge support from the Austrian Science Fund
(FWF) through projects F41 (SFB ViCoM), P27738-N28, and W1243 (Solids4Fun).
L.K. acknowledges support from the TU-D doctoral college (TU Wien).
M.A.L.M. acknowledges partial support from the German DFG through the
project MA6787/6-1.

\end{acknowledgments}

\bibliography{/planck/tran/divers/references}

\end{document}